\documentclass[aps,pra,twocolumn,superscriptaddress,showpacs,amsmath,amssymb]{revtex4-1}
\usepackage[dvips]{graphicx}
\usepackage{bm}

\newcommand{\vA}{{\bf A}}
\newcommand{\vp}{{\bf p}}
\newcommand{\vP}{{\bf P}}
\renewcommand{\vr}{{\bf r}}
\newcommand{\vR}{{\bf R}}
\newcommand{\vF}{{\bf F}}

\newcommand{\vq}{{\bf q}}
\newcommand{\vQ}{{\bf Q}}

\newcommand{\ve}{{\bf e}}

\newcommand{\cA}{{\cal A}}

\newcommand{\cE}{{\cal E}}
\newcommand{\cI}{{\cal I}}

\newcommand{\Ai}{{\rm Ai}}
\renewcommand{\Re}{{\rm Re}}
\renewcommand{\Im}{{\rm Im}}

\newcommand{\veps}{\hat{\boldsymbol{\epsilon}}}
\newcommand{\vkap}{\hat{\boldsymbol{\kappa}}}
\newcommand{\vt}{{\bf t}}
\newcommand{\vgamma}{\boldsymbol{\gamma}}
\newcommand{\vnu}{\boldsymbol{\nu}}

\newcommand{\ZETF}{Zh. Eksp. Teor. Fiz. }
\newcommand{\SPJ}{Sov. Phys. JETP }

\newcommand{\PRL}{Phys. Rev. Lett. }
\newcommand{\PR}{Phys. Rev. }
\newcommand{\jpb}{J. Phys. B }

\begin{document}

\title
{Analytic description of elastic electron-atom scattering
in an elliptically polarized laser field}

\author{A. V. Flegel}
\affiliation{Department of Physics and Astronomy, The University of Nebraska, Lincoln, NE 68588-0299}
\affiliation{Department of Computer Science, Voronezh State University, Voronezh 394006, Russia}
\author{M. V. Frolov}
\affiliation{Department of Physics, Voronezh State University, Voronezh 394006, Russia}
\author{N. L. Manakov}
\affiliation{Department of Physics, Voronezh State University, Voronezh 394006, Russia}
\author{Anthony F. Starace }
\affiliation{Department of Physics and Astronomy, The University of Nebraska, Lincoln, NE 68588-0299}
\author{A. N. Zheltukhin}
\affiliation{Department of Physics, Voronezh State University, Voronezh 394006, Russia}

 \date{\today}

\begin{abstract}
An analytic description of laser-assisted electron-atom scattering (LAES) in an elliptically
polarized field is presented using time-dependent effective range (TDER) theory to treat both
electron-laser and electron-atom interactions non-perturbatively. Closed-form formulas describing
plateau features in LAES spectra are derived quantum mechanically in the low-frequency limit. These
formulas provide an analytic explanation for key features of the LAES differential cross section.
For the low-energy region of the LAES spectrum, our result generalizes the Kroll-Watson formula to
the case of elliptic polarization. For the high-energy (rescattering) plateau in the LAES spectrum,
our result generalizes prior results for a linearly polarized field valid for the high-energy end
of the rescattering plateau [A.\,V. Flegel \textit{et al.}, J. Phys.~B \textbf{42}, 241002 (2009)]
and confirms the factorization of the LAES cross section into three factors: two field-free elastic
electron-atom scattering cross sections (with laser-modified momenta) and a laser field-dependent
factor (insensitive to the scattering potential) describing the laser-driven motion of the electron
in the elliptically polarized field. We present also approximate analytic expressions for the exact
TDER LAES amplitude that are valid over the entire rescattering plateau and reduce to the
three-factor form in the plateau cutoff region. The theory is illustrated for the cases of $e$-H
scattering in a CO$_2$-laser field and $e$-F scattering in a mid-infrared laser field of wavelength
$\lambda=3.5\,\mu$m, for which the analytic results are shown to be in good agreement with exact
numerical TDER results.
\end{abstract}


\maketitle

\section{Introduction}

The interaction of an intense laser field with atoms or molecules results in highly nonlinear
processes whose spectra are characterized by plateau-like structures, i.e. by a nearly constant
dependence of the cross sections on the number $n$ of absorbed photons over a wide interval of $n$.
These plateaus are well known for spectra of above-threshold ionization (ATI) and high-order
harmonic generation (HHG)~\cite{Salieres, Becker02, Ehlotzky1}. The rescattering
picture~\cite{Kuchiev, Shafer, Corkum} provides a transparent physical explanation for the
appearance of plateau structures: an intense oscillating laser field returns ionized electrons back
to the parent ion, whereupon they either gain additional energy from the laser field during
laser-assisted collisional events, thereby forming the high-energy plateau in ATI spectra, or
recombine with the parent ion, emitting high-order harmonic photons. High-energy plateaus
originating from laser-driven electron rescattering were predicted also for laser-assisted
radiative electron-ion recombination or attachment~\cite{Milo02, JETPL11} and laser-assisted
electron-atom scattering (LAES)~\cite{jetpl02, Milo04}. For \emph{laser-induced} bound-bound (as in
HHG) and bound-free (ATI) transitions, rescattering effects are suppressed for  an elliptically
polarized laser field and completely disappear for circular polarization. In contrast, for
\emph{laser-assisted} collisional processes (such as LAES) a rescattering plateau exists even for a
circularly polarized laser field~\cite{circ05} (cf.~also Ref.~\cite{Milo_el12}). The classical
rescattering scenario used to explain plateaus in LAES spectra for a linearly polarized field has
been justified by a quantum-mechanically derived analytic formula for the LAES differential cross
section~\cite{analit_laes}, which provides the rescattering correction to the well-known
Bunkin-Fedorov~\cite{BF} and Kroll-Watson~\cite{KW} results. This formula factorizes the LAES cross
section into the product of two field-free cross sections for elastic electron-atom scattering with
laser-modified momenta and a ``propagation'' factor (insensitive to atomic parameters) describing
the laser-driven motion of the electron along a closed classical trajectory. These three factors
provide closed-form quantum expressions for each of the three steps of classical rescattering
scenario for the LAES process.

Besides its fundamental interest for understanding better the physics of nonlinear phenomena,
factorization of the outcomes for nonlinear laser-atom processes in terms of laser-dependent
factors and factors describing the field-free atomic dynamics provides an efficient means for
retrieving these atomic factors from measured spectra of strong-field processes. At present, such
factorizations form the basis for HHG and ATI spectroscopies that allow the retrieval of the
photoionization cross sections for the outer electron shells of atoms or molecules (from HHG
spectra) (cf., e.g., Ref.~\cite{NatureTralerro}) and differential cross sections of elastic
electron scattering from the positive ion of a target (from ATI spectra) (cf., e.g.,
Refs.~\cite{Okunishi2008, Ray2008}). The factorization of HHG and ATI yields was first postulated
based on numerical solutions of the time-dependent Schr\"{o}dinger equation~\cite{MLZLPRL08}
(cf.~also the review~\cite{Lin_JPB2010}) and was then justified theoretically [within the
time-dependent effective range (TDER) theory~\cite{TDER2003, TDER2008}] for the case of a
monochromatic field in Refs.~\cite{JPB2009, FMSERSPRL09} for HHG and in Ref.~\cite{analit_atd} for
ATI, and for the case of a short laser pulse in Refs.~\cite{FMSVSPRA11} (for HHG)
and~\cite{OurPRL2012} (for ATI). We note that in all the aforementioned studies only linearly
polarized laser fields were considered, in which case the theoretical treatment is simplified (due
to the one-dimensional laser-driven propagation of the active electron along the direction of laser
polarization). However, although the driving laser ellipticity provides an additional control
parameter for intense laser-atom interactions, at present there does not exist a convincing
justification for the factorization of the rates or cross sections of nonlinear phenomena in an
elliptically polarized field, neither for laser-induced nor for laser-assisted processes.

In this paper we show analytically that the LAES cross section in the region of the rescattering
plateau cutoff may be expressed in factorized form (as the product of three factors) for the
general case of an elliptically polarized laser field. This result generalizes that for the case of
linear polarization~\cite{analit_laes} and presents a rare example of a strong field process whose
yield may be factorized for the case of a nonzero driving laser ellipticity. The results presented
are obtained taking into account the rescattering effects non-perturbatively within the TDER theory
for collision problems~\cite{jetpl08} as reformulated for the case of LAES in a low-frequency,
elliptically polarized field. Based on a detailed analysis of the two-dimensional closed classical
trajectories of an electron in the laser polarization plane, we have obtained also an analytic
estimate for the (non-factorized) LAES amplitude that describes the entire energy region of the
rescattering plateau. Our analytic results are in good agreement with exact numerical TDER results.

The paper is organized as follows. In Sec.~II we provide the basic results of the TDER theory for
the scattering state of an electron as well as for the LAES amplitude in an elliptically polarized
laser field. In Sec.~III we develop a low-frequency expansion for  the key ingredient of TDER
theory: the periodic function of time, $f_{\bf p}(t)$, that enters the TDER result for the
scattering state. This expansion allows one to approximate the scattering state as a sum of two
terms: a zero-order (``Kroll-Watson'') term and a rescattering correction, which is responsible for
the high-energy plateau in the LAES spectrum. The low-energy part of the LAES spectrum, described
by the Kroll-Watson term in the LAES amplitude, is considered in Sec.~IV, while in Sec.~V we
provide a detailed analysis of the LAES amplitude in the rescattering approximation, i.e.,
including the rescattering correction.  In Sec.~VI we present the factorized (three-factor) form
for the LAES cross section in the rescattering approximation, compare the LAES spectra in this
approximation with exact TDER results, and discuss the influence of the laser ellipticity on key
features of LAES spectra. Some conclusions and perspectives for further use of the TDER theory for
description of LAES in an elliptically polarized field are discussed briefly in Sec.~VII. Finally,
in two Appendices we present an alternative representation for the TDER LAES amplitude that we use
for the exact numerical calculations within the TDER theory (Appendix A) and a brief description of
the uniform asymptotic approximation of an integral involving a highly-oscillatory function
(Appendix B).

\section{Basic equations of the TDER theory for LAES}
\subsection{Formulation of the problem}
We consider the scattering of an incoming electron having momentum $\vp$ and kinetic energy
$E=p^2/(2m)$ on a target atom in the presence of a long laser pulse approximated by a
monochromatic, elliptically polarized plane wave having intensity $I$ and frequency $\omega$. We
assume that both the electron energy $E$ and the laser photon energy $\hbar\omega$ are small
compared to atomic excitation energies and that the laser parameters $I$ and $\omega$ are such that
laser excitation/ionization of atomic electrons is negligible. Under these assumptions, the
electron-atom interaction can be approximated by a short-range potential $U(r)$ (that vanishes for
$r \gtrsim r_c$). Thus, the LAES process can be described as potential (elastic) electron
scattering accompanied by absorption or emission of $n$ laser photons (with
$n_{\min}=-[E/(\hbar\omega)]$, where $[x]$ is the integer part of $x$). Thus, the momentum (or
energy) spectra of the scattered electrons (the LAES spectra) are characterized by momenta $\vp_n$
and energies $E_n=p_n^2/(2m)=E+n\hbar\omega$.

For the electron-laser interaction, we use the dipole approximation in the length gauge,
\begin{equation}
V(\vr,t)=-e\vr\cdot\vF(t),
\end{equation}
were $\vF(t)$ is the electric vector of the laser field,
\begin{equation}
\label{laser:F} \vF(t) = F\Re\big(\ve e^{-i\omega t}\big), \quad \ve\cdot\ve^*=1.
\end{equation}
The complex polarization unit vector $\ve$ in Eq.~(\ref{laser:F}) is parameterized as
\begin{equation}
\label{laser:pol}
\ve = \frac{\veps + i\eta[\vkap \times \veps]}
{\sqrt{1+\eta^2}},
\quad
-1\leq\eta\leq 1,
\end{equation}
where $\veps$ is a unit vector along the major axis of the polarization ellipse, the unit vector
$\vkap$ defines the laser propagation direction, and $\eta$ is the ellipticity. With
the definition (\ref{laser:pol}), the laser intensity does not depend on $\eta$: $I=cF^2/(8\pi)$.
Along with $\eta$, the degrees of linear ($\ell$) and circular ($\xi$) polarization are convenient
parameters for describing an elliptically polarized field:
\begin{equation}
\ell=\ve\cdot\ve = \frac{1-\eta^2}{1+\eta^2},\quad
\xi=i\vkap\cdot[\ve\times\ve^*] = \frac{2\eta}{1+\eta^2}.
\label{ell-xi}
\end{equation}
Note that the scalar product of the polarization vector $\ve$ with a unit vector $\mathbf{u}$, defined
by the two spherical angles, $\theta_{\mathbf{u}}$ and $\phi_{\mathbf{u}}$, as $\mathbf{u} =
(\veps\cos\phi_{\mathbf{u}}+ [\vkap \times\veps]\sin\phi_{\mathbf{u}})\sin\theta_{\mathbf{u}}
+\vkap\cos\theta_{\mathbf{u}}$, is complex and can be parametrized as
\begin{eqnarray}
& & \mathbf{u}\cdot\ve = |\mathbf{u}\cdot\ve |e^{i\varphi_{\mathbf{u}}},\quad
\varphi_{\mathbf{u}}\equiv \arg(\mathbf{u}\cdot\ve),\nonumber \\
& & |\mathbf{u}\cdot\ve | = \sin\theta_{\mathbf{u}}\sqrt{(1+\ell\cos 2\phi_{\mathbf{u}})/2},
\label{scal-prod} \\
& & \tan\varphi_{\mathbf{u}} = \eta\tan\phi_{\mathbf{u}}.
\nonumber
\end{eqnarray}

For an analytic non-perturbative account of both the electron-laser and the electron-atom
interactions in electron scattering assisted by a low-frequency elliptically polarized laser field,
we adapt the TDER theory~\cite{jetpl08} for LAES to the case of a low-frequency field. The atomic
potential $U(r)$ is assumed to support a single (negative ion) weakly-bound state $\psi_{\kappa l
m_l}(\vr)$ with energy $E_0=-\hbar^2\kappa^2/(2m)$ $(\kappa r_c \ll 1)$ and angular momentum $l$.
In particular, $l=0$ corresponds to electron scattering from hydrogen or an alkali atom, and $l=1$
corresponds to a halogen atom target.

The key idea of the TDER theory is the same as in effective range theory for two stationary
potentials, $U(r)$ and $V(\vr)$, which exert their influence on the electron predominantly in two
essentially non-overlapping coordinate ranges~\cite{TMF85}: $U(r)$ is important primarily for
$r\lesssim r_c$, while a long-range, external-field potential $V(\vr)$ is important primarily for
$r\gg \kappa^{-1}$. Thus, in the region $r_c\lesssim r\ll \kappa^{-1}$, the low-energy electron may
be considered as virtually free. In this case, as in effective range theory for low-energy electron
scattering~\cite{LL}, only a single parameter, the $l$-wave scattering phase $\delta_l$ for the
potential $U(r)$, determines the $l$-wave component of the exact scattering state $\psi_\vp(\vr)$
in the region $r_c\lesssim r\ll \min(\kappa^{-1},k^{-1})$ ($k=\sqrt{2mE}/\hbar=p/\hbar$):
\begin{equation}
\label{BC:free} \int Y^*_{lm_l}(\hat\vr)\psi_\vp(\vr)d\Omega_\vr \sim r^{-l-1} + \cdots +
B_l(E)(r^l + \cdots),
\end{equation}
where the factor $B_l(E)$ involves the phase shift $\delta_l(k)$ and can be approximated by two
fundamental parameters of the effective range theory: the scattering length ($a_l$) and the
effective range ($r_l$):
\begin{eqnarray}
& & (2l-1)!!(2l+1)!!B_l(E) \equiv k^{2l+1}\cot\delta_l(k) \nonumber \\
& & = -a_l^{-1} + r_lk^2/2.
\label{B_l}
\end{eqnarray}
The boundary condition (\ref{BC:free}) for $\psi_\vp(\vr)$ at small $r$ is the key equation that
allows one to represent the scattering state $\psi_\vp(\vr)$ outside the potential $U(r)$ (i.e.,
for $r\gtrsim r_c$) in terms of the two parameters of the problem, $a_l$ and $r_l$, independent of
the shape of $U(r)$.

\subsection{Scattering state of an electron in TDER theory}

We seek the laser-dressed scattering state, $\Psi_\vp(\vr,t)$, of an electron in the LAES process
using the Floquet or quasienergy state (QES) representation (cf., e.g., Ref.~\cite{PhRep}):
\begin{equation}
\label{qes} \Psi_\vp(\vr,t)=e^{-i\epsilon t/\hbar}\Phi_\vp(\vr,t),\;
\Phi_\vp(\vr,t)=\Phi_\vp(\vr,t+2\pi/\omega),
\end{equation}
where $\epsilon=E+u_p$ is the quasienergy and $u_p=e^2F^2/(4m\omega^2)$ is the ponderomotive (or
quiver) energy. The QES wave function $\Phi_\vp(\vr,t)$ is a periodic solution of the
time-dependent Schr\"{o}dinger equation:
\begin{equation}\label{TDSE}
\left(i\hbar\frac{\partial }{\partial t}+\epsilon +\frac{\hbar^2}{2m}\Delta -U(r) - V({\bf r},t)
\right)\Phi_\vp(\vr,t)=0.
\end{equation}
Owing to the time dependence of $\Phi_\vp(\vr,t)$, the boundary condition for the $l$-wave
component of $\Phi_\vp(\vr,t)$ at small $r\gtrsim r_c$ must be modified compared to
Eq.~(\ref{BC:free}) by introducing some time-periodic functions (as was done similarly in
TDER theory for bound states in an elliptically polarized field~\cite{TDER2003, TDER2008}). Since $V({\bf r},t)$ lacks axial symmetry in the case of an elliptically polarized field, the $l$-wave
component of $\Phi_\vp(\vr,t)$ depends in general  on the angular momentum projection $m_l$.
However, for small $r\gtrsim r_c$ the potentials $U(r)$ and $V({\bf r},t)$ can be neglected in Eq.~(\ref{TDSE}), so that the $l$-wave component of any time-periodic solution of Eq.~\eqref{TDSE} is independent of $m_l$ and may be written as:
\begin{eqnarray}
&&\int Y_{l m_l}^*(\hat\vr)\Phi_\vp(\vr,t)d\Omega_\vr   \nonumber \\
&&=\sum_{k}[a_k j_l(\varkappa_k r)+b_k y_l(\varkappa_kr)]e^{-ik\omega t}, \label{philml}
\end{eqnarray}
where $\varkappa_k=\sqrt{2m(\epsilon+k\hbar\omega)}/\hbar$, $j_l$ and $y_l$ are the regular and irregular spherical Bessel functions (behaving respectively as $\sim r^l$ and $\sim
r^{-l-1}$ as $r \rightarrow 0$), and $a_k$ and $b_k$ are constants. Replacing $j_l(\varkappa_k r)$ and $y_l(\varkappa_k r)$ in Eq.~(\ref{philml}) by their expansions for $\varkappa_k r\ll 1$, defining the factor $B_l(\epsilon + k\hbar\omega)$ as proportional to the coefficient ratio $a_k/b_k$, and introducing coefficients $f_k^{(lm_l)}$, in which the index $m_l$ labels the angular momentum projection onto $Y_{l m_l}$ on the left of Eq.~\eqref{philml}, we obtain a generalization of the boundary condition (\ref{BC:free}) for a time-dependent interaction $V(\vr,t)$:
\begin{eqnarray}
\nonumber &&\int Y^*_{lm_l}(\hat\vr)\Phi_\vp(\vr,t)d\Omega_\vr
\sim \sum_k [r^{-l-1} + \cdots \\
&& \nonumber +\, B_l(\epsilon + k\hbar\omega)(r^l + \cdots)] f_k^{(lm_l)}e^{-ik\omega t}\\
&&=\, \left[r^{-l-1}+\cdots+B_l(\epsilon)(r^l+\cdots)\right]f_\vp^{(lm_l)}(t)\nonumber\\
&&+\,i(r^l+\cdots)\frac{(2l+1)}{\left[(2l+1)!!\right]^2}\frac{r_lm}{\hbar}\frac{d}{dt}f^{(lm_l)}_\vp(t),
\label{BC:qes}
\end{eqnarray}
where the effective range parametrization (\ref{B_l}) for $B_l(\epsilon + k\hbar\omega)$ was substituted on the left of the equality in Eq.~\eqref{BC:qes} in order to obtain the final result summed over $k$ on the right in terms of the time-periodic function
\begin{equation}\label{fplml}
f_\vp^{(lm_l)}(t) = \sum_kf_k^{(lm_l)} e^{-ik\omega t}.
\end{equation}

The desired solution of the exact equation (\ref{TDSE}) for the scattering states has the following
general form:
\begin{equation}
\label{scatt.state} \Phi_\vp(\vr,t) = \chi_\vp(\vr,t) + \Phi_\vp^{(sc)}(\vr,t),
\end{equation}
where the ``scattered wave'' $\Phi_\vp^{(sc)}(\vr,t)$ is an outgoing wave at $r\to\infty$, while the ``incident wave'' $\chi_\vp(\vr,t)$ is the QES wave function of a free electron with momentum $\vp$ in the laser field (i.e., the time-periodic part of a Volkov wave function),
\begin{equation}
\chi_\vp(\vr,t) = e^{i[\vr\cdot\vP(t) - S_\vp(t)]/\hbar},
\label{Volkov}
\end{equation}
where
\begin{eqnarray}
\nonumber
S_\vp(t) & = & \int^t [\vP^2(\tau)/(2m) - \epsilon]d\tau \\
& = & -\vp\cdot\frac{e\vF(t)}{m\omega^2} +
\int^t\left[\frac{e^2\vA^2(\tau)}{2mc^2}-u_p\right]d\tau, \label{Sp}
\end{eqnarray}
and $\vP(t) = \vp - (e/c)\vA(t)$ is the electron's kinetic momentum in the laser field $\vF(t)$ with vector potential $\vA(t)$, where $\vF(t)=-c^{-1}d\vA(t)/dt$.

According to the TDER theory~\cite{jetpl08}, the function $\Phi_\vp^{(sc)}(\vr,t)$ in the outer
region, $r\gtrsim r_c$ [in which the potential $U(r)$ vanishes], can be expressed in terms of the
retarded Green's function $G(\vr,t;\vr,t')$ of a free electron in the laser field $\vF(t)$ and
involves the function $f^{(lm_l)}_\vp(t)$ in the boundary condition (\ref{BC:qes}). [Indeed,
upon neglecting $U(r)$, any solution of Eq.~(\ref{TDSE}) can be represented as a wave packet composed
of wave functions for a free electron in the field $\vF(t)$.] For $G(\vr,t;\vr,t')$ we use the
well-known Feynman form:
\begin{eqnarray}
\nonumber
G(\vr,t;\vr',t') &=& -\theta(t-t')\frac{i}{\hbar}\left[ \frac{m}{2\pi i\hbar(t-t')} \right]^{3/2}\\
&\times & \exp[iS(\vr,t;\vr',t')/\hbar],
\label{Green}
\end{eqnarray}
where $\theta(x)$ is Heaviside function and $S$ is the classical action for an electron in the laser
field $\vF(t)$:
\begin{eqnarray}
&&\!\! S(\vr,t;\vr',t') = \frac{m}{2 (t-t')}\left(\vr - \vr' + \frac{e}{m\omega^2}[\vF(t)-\vF(t')] \right)^2
\nonumber
\\&&\quad - \frac{e^2}{2mc^2}\int_{t'}^{t}\vA^2(\tau)d\tau - \frac{e}{c}\left[\vr\cdot\vA(t)-\vr'\cdot\vA(t')\right].
\label{action}
\end{eqnarray}

The behavior of $\Phi_\vp(\vr,t)$ as $r\to 0$  required by the condition (\ref{BC:qes}) [namely,
the $l$-wave component of $\Phi_\vp(\vr,t)$ should involve a singular term $\sim
r^{-l-1}Y_{lm_l}(\hat\vr)$] may be ensured by $l$-fold differentiation of $G(\vr,t;\vr,t')$ over
$\vr'$ followed by the substitution $\vr'=0$. [From the explicit form (\ref{Green}) of $G$, such
differentiation does not change the asymptotic behavior of $\Phi_\vp(\vr,t)$ for $r\to\infty$.] As
a result, in a way similar to that for the TDER treatment of a quasistationary quasienergy state
with an initial angular momentum $l$~\cite{TDER2003, TDER2008}, the general TDER expression for
$\Phi_\vp^{(sc)}(\vr,t)$ can be written as follows~\cite{jetpl08}:
\begin{eqnarray}
\label{wf:out:l} \nonumber \Phi_\vp^{(sc)}(\vr,t) & = &
-\frac{2\pi\hbar^2}{m\kappa^{1+l}}\sum_{\mu=-l}^{l}
\int_{-\infty}^t dt'\,e^{i\epsilon(t-t')/\hbar}f_\vp^{(l\mu)}(t')\\
& \times & \mathcal{Y}_{l\mu}(\nabla_{\vr'})G(\vr,t;\vr',t') \big|_{\vr'=0},
\end{eqnarray}
where the differential operator $\mathcal{Y}_{l\mu}(\nabla_{\vr})$ is obtained from the solid
harmonic $\mathcal{Y}_{l\mu}(\vr)$ $[\equiv r^lY_{l\mu}(\hat\vr)]$ by the substitution
$\vr\to\nabla_\vr$. Equations for the unknown functions $f_\vp^{(l\mu)}(t)$ complete the
construction of the scattering state $ \Phi_\vp(\vr,t)$ in TDER theory
[cf.~Eqs.~(\ref{scatt.state}) and (\ref{wf:out:l})]. These equations are obtained by matching the
$l$-wave components of $ \Phi_\vp(\vr,t)$ [which are different for different values of $m_l$, as
noted above and as is clear from the explicit representation (\ref{wf:out:l}) for
$\Phi_\vp^{(sc)}(\vr,t)$] at small $r$  to the prescribed boundary condition~\eqref{BC:qes}. Due to
the term $\chi_\vp(\vr,t)$ in Eq.~(\ref{scatt.state}), the resulting equations comprise a system of
$2l+1$ coupled inhomogeneous integro-differential equations for the functions $f_\vp^{(lm_l)}(t)$,
with $m_l =-l, \cdots, l$. Because the derivation and analysis of these equations involve the same
steps for both $l>0$ and $l=0$ (differing only in the complexity of the analytical
transformations),  for greater clarity, in the rest of this paper we provide analytical derivations
only for the case of $l=0$ (``$s$-wave scattering''). (For an analytical treatment of a similar,
though homogeneous, system of equations in TDER theory for bound states with $l>0$, see
Refs.~\cite{TDER2003, TDER2008}.)

\subsection{Exact TDER LAES amplitude and differential cross section for $s$-wave scattering}

If the potential $U(r)$ supports only a single weakly-bound $s$-state so that only the phase shift
$\delta_0(k)$ is non-zero, then Eqs.~(\ref{BC:qes})  and (\ref{wf:out:l}) simplify as follows:
\begin{eqnarray}
&& \Phi_\vp^{(sc)}(\vr,t)  =  -\frac{2\pi\hbar^2}{m\kappa}
\int_{-\infty}^t dt'\,e^{i\epsilon(t-t')/\hbar}f_\vp(t') \nonumber \\
&&\times  G(\vr,t;0,t'), \label{wf:out}\\
&& \Phi_\vp(\vr,t) \sim \left(\frac{1}{r} + B_0(\epsilon)\right)f_\vp(t) +
i\frac{r_0m}{\hbar}\frac{d}{dt}f_\vp(t), \label{BC:S}
\end{eqnarray}
where $f_\vp(t) \equiv f_\vp^{(00)}(t)$ and
\begin{equation}
B_0(\epsilon) = - a_0^{-1} + r_0m\epsilon/\hbar^2. \label{Bzero}
\end{equation}
To match the function $\Phi_\vp(\vr,t)$ [cf.~Eqs.~(\ref{scatt.state}),~\eqref{wf:out}] to the $r
\rightarrow 0$ boundary condition (\ref{BC:S}), we extract from the integrand in Eq.~(\ref{wf:out})
a term proportional to the field-free Green's function $G_0(\vr,t;0,t')$ [given by
Eq.~(\ref{Green}) with $\vF(t)=0$]:
\begin{eqnarray}
\nonumber
&& \Phi_\vp^{(sc)}(\vr,t) = -\frac{2\pi\hbar^2}{m\kappa}
\int_{-\infty}^t dt'\big[\,e^{i\epsilon(t-t')/\hbar}f_\vp(t')\\
&& \times G(\vr,t;0,t')
-f_\vp(t)G_0(\vr,t;0,t')\big] + \frac{1}{\kappa r}f_\vp(t).\;\;
\label{wf:out2}
\end{eqnarray}
The integral in Eq.~(\ref{wf:out2}) is now regular at $\vr=0$. Setting then $\vr=0$ in
$\chi_\vp(\vr, t)$, comparing the result for $\Phi_\vp(\vr,t)$ at small $r$ with Eq.~(\ref{BC:S}),
and introducing the dimensionless time $\tau=\omega t$, we obtain an inhomogeneous
integro-differential equation for $f_\vp(\tau)\equiv f_\vp(t=\tau/\omega)$:
\begin{eqnarray}
&& B_0(\epsilon)f_\vp(\tau) +i\frac{r_0m\omega}{\hbar}\frac{d}{d\tau}f_\vp(\tau)
\nonumber \\
&& = \kappa e^{-iS_\vp(\tau)/\hbar} + \cI[f_\vp(\tau)],
\label{intEq:f}
\\
&&\cI[f_\vp(\tau)] = \sqrt{\frac{m\omega}{2\pi i\hbar}}
\int_0^{\infty}\frac{dx}{x^{3/2}}\big[
e^{(i/\hbar)[\epsilon x/\omega + S(\tau,\tau-x)]} \nonumber \\
&&\phantom{\cI[f_\vp(\tau)]} \times f_\vp(\tau-x) - f_\vp(\tau)\big],
\label{Int_term}
\end{eqnarray}
where $S_\vp(\tau)\equiv S_\vp(t=\tau/\omega)$, $S(\tau,\tau')\equiv
S(\vr=0,t=\tau/\omega;\vr'=0,t'=\tau'/\omega)$.

As is usual, the LAES amplitude $\cA_n(\vp,\vp_n)$ is determined by the asymptotic behavior of the wave function $\Phi_\vp^{(sc)}(\vr,t)$ in Eq.~(\ref{wf:out:l}) as $r\to\infty$. For $s$-wave scattering, this behavior has the form:
\begin{eqnarray}
&&\Phi_\vp^{(sc)}(\vr,t)\Big\vert_{\kappa r\gg 1} \backsimeq
e^{-i\phi(\vr,t)/\hbar} \nonumber \\
&&\times\sum_{n\geq n_{\min}}\cA_n(\vp,\vp_n)
\frac{e^{ip_n|\vR(\vr,t)|/\hbar - in\omega t}}{|\vR(\vr,t)|},
\label{wf:asimptot}
\end{eqnarray}
where
\begin{eqnarray*}
&&\phi(\vr,t)=\frac{e}{c}\vr\cdot\vA(t) + \int^t\left(\frac{e^2A^2(\tau)}{2mc^2}-u_p\right)d\tau,\\
&&\vR(\vr,t)=\vr + \frac{e}{m\omega^2}\vF(t),
\end{eqnarray*}
and the summation over $n$ involves all open channels with exchange of $n$ photons, for which $E_n
= E+n\hbar\omega > 0$. The LAES amplitude $\cA_n(\vp,\vp_n)$ may be expressed in
terms of $f_\vp(\tau)$,
\begin{equation}
\label{amplitude:f}
\cA_n(\vp,\vp_n) = \frac{1}{2\pi\kappa}\int_0^{2\pi}e^{in\tau + iS_{\vp_n}(\tau)/\hbar}f_\vp(\tau)d\tau,
\end{equation}
and the differential LAES cross section is given by
\begin{equation}
\label{cross}
\frac{d\sigma_n(\vp,\vp_n)}{d\Omega_{\vp_n}} = \frac{p_n}{p}\left|\cA_n(\vp,\vp_n)\right|^2.
\end{equation}

For $\vF(t) =0$, the function $f_\vp(\vF(t) =0;\tau)\equiv f_0(p)$ reduces to the amplitude
$\cA(p)$ for field-free $s$-wave elastic electron scattering on the potential $U(r)$ in the
effective range approximation (in which $k=p/\hbar$),
\begin{equation}
\label{ampl_el:s} f_0(p)=\kappa \cA(p), \quad \cA(p) = \frac{1}{- a_0^{-1} + r_0 k^2/2 - ik }.
\end{equation}
For $\vF(t) \neq 0$, the function $f_\vp(\tau)$ is a key object of TDER theory, since it contains
complete information on the modification of the electron-atom interaction by an elliptically
polarized laser field in all LAES channels. Numerical evaluation of $f_\vp(\tau)$ is done most
easily by converting the integro-differential Eq.~(\ref{intEq:f}) to a set of inhomogeneous linear
algebraic equations for the Fourier-coefficients $f_k$ of $f_\vp(\tau)$ [cf.~Eq.~(\ref{s:system})
in Appendix~\ref{app1}]. The LAES amplitude $\cA_n(\vp,\vp_n)$ is then expressed in terms of $f_k$
and generalized Bessel functions [cf.~Eq.~(\ref{s:amplitude})]. As follows from the boundary
condition (\ref{BC:qes}) that determines the QES $\Phi_\vp(\vr,t)$ at small $r$ [where the
potential $U(r)$ is most important], physically, the coefficients $f_k$ govern the population of
QES harmonics of the scattering state $\Psi_\vp(\vr,t)$ with energies $\epsilon +k\hbar\omega$ that
arise as a result of atomic-potential-mediated exchange of $k$ photons between the electron and the
laser field at small $r$.

For $s$-wave scattering, the numerical results in this paper, referred to as ``exact TDER
results,'' are obtained by numerical solution of Eq.~(\ref{s:system}), followed by evaluation of
the amplitude $\cA_n(\vp,\vp_n)$ according to Eq.~(\ref{s:amplitude}). For $p$-wave scattering
($l=1$), the Fourier coefficients $f_k^{(1m)}$ (where $m=0,\,\pm1$) of the periodic
function~\eqref{fplml} satisfy the system of Eqs.~\eqref{p:eq1}, \eqref{p:eq2}, while the LAES
amplitude is given by Eq.~\eqref{p:amplitude}.

An analytic evaluation of the LAES amplitude $\cA_n(\vp,\vp_n)$ can be performed in the
low-frequency limit, in which case the low-frequency expansion for the solution $f_\vp(\tau)$ of
Eq.~(\ref{intEq:f}) can be obtained.

\section{Low-frequency expansion of $f_\vp(\tau)$ \label{low-frequency}}

Because the classical actions $S_\vp(\tau)$ and $S(\tau,\tau')$ in the inhomogeneous and integral terms of Eq.~(\ref{intEq:f}) oscillate with large amplitudes ($\sim u_p/\omega$) for the case of an intense low-frequency field $\vF(t)$, we seek the solution $f_\vp(\tau)$ of Eq.~(\ref{intEq:f}) in the following form:
\begin{equation}
\label{f:gen}
f_\vp(\tau)=g_\vp(\tau)e^{-i\int^{\tau}\! d\tau'[\cE(\tau')-\epsilon]/(\hbar\omega)},
\end{equation}
where $g_\vp(\tau)$ and $\cE(\tau)$ are smooth functions satisfying respectively the requirements that $|dg_\vp/d\tau| \ll u_p/(\hbar\omega)$ and that the upper bound of $\cE(\tau)$ is of the order of $u_p$.

Before proceeding  with an iterative solution of Eq.~(\ref{intEq:f}), we analyze first the low-frequency limit of the integral term $\cI[f_\vp(\tau)]$ defined in Eq.~(\ref{Int_term}). For $u_p\gg\hbar\omega$, the dominant contribution to the integral~(\ref{Int_term}) comes from the neighborhood of the singular point $x=0$, while the contribution from the domain $x >0$ can be evaluated using the saddle point method. Thus we approximate the integral $\cI$ [after substituting there Eq.~(\ref{f:gen})] as a sum:
\begin{equation}
\label{Int_approx}
\cI[f_\vp(\tau)]\approx \cI^{(0)}[f_\vp(\tau)] + \sum_s\cI_s[f_\vp(\tau)],
\end{equation}
where integrals $\cI^{(0)}$ and $\cI_s$ are evaluated respectively over the vicinity of $x=0$ and at the saddle points $x=x_s > 0$. In order to evaluate the term $\cI^{(0)}$, we neglect the action $S(\tau,\tau-x)$ (which is of order $x^3$ when $x\to0$) in the integrand of Eq.~(\ref{Int_term}) and approximate the function $f_\vp(\tau-x)$ for $x\ll 1$ as
\begin{equation*}
f_\vp(\tau-x)\approx f_\vp(\tau)e^{ix[\cE(\tau)-\epsilon]/(\hbar\omega)}.
\end{equation*}
We thus obtain for $\cI^{(0)}$ the following result:
\begin{eqnarray}
\nonumber
{\cal I}^{(0)}[f_\vp(\tau)]\! &=&\! f_\vp(\tau)\sqrt{\frac{m\omega}{2\pi i\hbar}}
\int_0^{\infty}\frac{dx}{x^{3/2}}\left(
e^{ix\cE(\tau)/(\hbar\omega)}-1\right)\\
&=& if_\vp(\tau)\sqrt{2m\cE(\tau)}/\hbar.
\label{Int_x0}
\end{eqnarray}
The result for $\cI_s$ is obtained by substituting Eq.~(\ref{f:gen}) into Eq.~(\ref{Int_term}), followed by evaluation of $\cI$ by the saddle point method:
\begin{equation}
{\cal I}_s[f_\vp(\tau)] = f_\vp(\tau-x_s)
\frac{e^{(i/\hbar)[\epsilon x_s/\omega + S(\tau,\tau-x_s)]}}{\alpha_0\big[x_s^3\beta(\tau,x_s)\big]^{1/2}},
\label{Int_xs}
\end{equation}
where we have introduced the dimensionless function $\beta(\tau,x)$,
\begin{equation}
\label{beta:gen}
\beta(\tau,x)=\frac{1}{4u_p}\frac{\partial}{\partial \tau'}\Big[\omega\,
\frac{\partial S(\tau,\tau')}{\partial \tau'}
-\cE(\tau')\Big]\Big|_{\tau'=\tau-x},
\end{equation}
and the quiver radius, $\alpha_0=|e|F/(m\omega^2)$,  for free-electron
oscillations in the field $\vF(t)$. The saddle points $x_s$ are solutions of the equation:
\begin{equation}
\label{saddle:gen}
\omega\,\frac{\partial}{\partial x}S(\tau,\tau-x) +
\cE(\tau-x) = 0.
\end{equation}

The results (\ref{Int_x0}) and (\ref{Int_xs}) for the integral terms $\cI^{(0)}$ and $\cI_s$ allow
us to develop an iterative procedure for the solution of Eq.~\eqref{intEq:f} for $f_\vp(\tau)$ in
the low-frequency limit. To do that,  we note that the saddle point contributions, $\cI_s$, to the
integral $\cI$ in the approximation (\ref{Int_approx}) are proportional to the dimensional
parameter $\alpha_0^{-1}=m\omega^2/(|e|F)$, while $\cI^{(0)}$ is proportional to
$\sqrt{2m\cE(\tau)}/\hbar$, where $\cE(\tau)\sim u_p$. Thus, the ratio of terms $\cI_s$ to
$\cI^{(0)}$ is determined by a dimensionless factor $\sim \hbar\omega/u_p$. Therefore, the
iterative account of terms  $\cI_s$ (which, as we will show below, describe the rescattering
effects in LAES) is valid at the condition
\begin{equation}
\label{parameter}
\frac{\hbar\omega}{u_p} \ll 1.
\end{equation}
It is worthwhile to emphasize that, besides the frequency, the condition (\ref{parameter}) involves
also the field amplitude $F$, so that the low-frequency expansion for the QES $\Phi_{\bf p}({\bf
r},t)$ can be called also a ``strong-field'' expansion, since already for ${\hbar\omega}/{u_p}
\lesssim 1$ the perturbation theory (in laser-atom interaction) for $\Phi_{\bf p}({\bf r},t)$
becomes divergent~\cite{PT}.

\subsection{The zero-order approximation for $f_\vp(\tau)$}

To obtain the zero-order approximation in the parameter $\hbar\omega/u_p$ for the function $f_\vp(\tau)$ [$f_\vp(\tau) \approx f_\vp^{(0)}(\tau)$], we note that the strongly-oscillating exponential factor in Eq.~(\ref{f:gen}) is determined by the inhomogeneous term of Eq.~(\ref{intEq:f}) [taking into account Eq.~\eqref{Sp}], so that
\begin{eqnarray}
&& \cE(\tau)=\cE^{(0)}(\tau)=\frac{\vP^2(\tau)}{2m},
\label{cE0} \\
&& f_\vp^{(0)}(\tau)=g_\vp^{(0)}(\tau)e^{-iS_\vp(\tau)/\hbar}.
\label{f0}
\end{eqnarray}
The pre-exponential factor $g^{(0)}_\vp(\tau)$ can be obtained from Eq.~(\ref{intEq:f}) after
substituting there $f_\vp(\tau) \to f_\vp^{(0)}(\tau)$, omitting then the differential term
($\sim\omega\, dg_\vp^{(0)}/d\tau$) and the saddle point contributions
to the integral (\ref{Int_term}) [retaining only the first term in Eq.~\eqref{Int_approx}, given by Eq.~\eqref{Int_x0}]. The result for $g_\vp^{(0)}(\tau)$ is thus
\begin{equation}
\label{g0}
g_\vp^{(0)}(\tau) = \frac{\kappa}{B_0[\cE^{(0)}(\tau)]-iP(\tau)/\hbar}
=\kappa\cA[P(\tau)],
\end{equation}
where the second equality, obtained using Eq.~\eqref{Bzero}, gives the amplitude $\cA[P(\tau)]$ [cf. Eq.~(\ref{ampl_el:s})] for laser-free elastic electron-atom
$s$-wave scattering in the effective range approximation, as a function of the time-dependent kinetic momentum $P(\tau)$.

\subsection{The first-order (rescattering) correction to $f_\vp^{(0)}(\tau)$}

The first-order iterative correction $f_\vp^{(1)}$ to the zero-order result
$f_\vp^{(0)}$ satisfies the equation obtained by substituting
\begin{equation} \label{f01}
f_\vp(\tau)=f_\vp^{(0)}(\tau)+f_\vp^{(1)}(\tau)
\end{equation}
into Eq.~(\ref{intEq:f}):
\begin{eqnarray}
&& \bigg(B_0[\cE^{(1)}(\tau)]-\frac{i}{\hbar}\sqrt{2m\cE^{(1)}(\tau)}\bigg)
f_\vp^{(1)}(\tau) \nonumber \\
&& =\sum_{s}\cI_s[f_\vp^{(0)}(\tau)],
\label{intEq:f1}
\end{eqnarray}
where $f_\vp^{(1)}$ is taken in the form (\ref{f:gen}) with $\cE(\tau)=\cE^{(1)}(\tau)$ and $g_\vp(\tau)=g_\vp^{(1)}(\tau)$. In deriving Eq.~(\ref{intEq:f1}), the differential term in Eq.~(\ref{intEq:f}) is evaluated as follows: we neglected the terms involving $dg_\vp^{(0)}/d\tau$ and $dg_\vp^{(1)}/d\tau$ and combined the result of taking the derivative of the exponential [see Eq.~\eqref{f:gen}] with the term involving $B_0(\epsilon)$ [see Eq.~\eqref{Bzero}] to obtain $B_0[\cE^{(1)}(\tau)]$. Also, we used Eq.~(\ref{Int_x0}) for $\cI^{(0)}[f_\vp^{(1)}(\tau)]$. The explicit form for $\cI_s[f_\vp^{(0)}(\tau)]$ follows from Eqs.~(\ref{Int_xs}) and (\ref{beta:gen}) taking into account Eqs.~(\ref{cE0}) and (\ref{f0}):
\begin{equation}
{\cal I}_s[f_\vp^{(0)}(\tau)] = g_\vp^{(0)}(\tau-x_s)
\frac{e^{i\varphi(\tau,x_s) - iS_\vp(\tau)/\hbar}}
{\alpha_0\big[x_s^3\beta(\tau,x_s)\big]^{1/2}},
\label{Int1_xs}
\end{equation}
where
\begin{eqnarray}
&&\varphi(\tau,x) = \frac{x[\vp-\vq(\tau, \tau - x)]^2}{2m\hbar\omega},
\label{Int:phase}
\\
&&\beta(\tau,x) =  \frac{\omega^2}{e^2F^2}
\Big\{\frac{e}{\omega}\vF(\tau-x)\cdot[\vq(\tau,\tau-x)-\vp]
\nonumber \\
&&\phantom{\beta(\tau,x)} + \vQ^2(\tau-x,\tau)/x\Big\}, \label{R:beta}
\\
&&\vQ(\tau,\tau') = \vq(\tau,\tau') -\frac{e}{c}\vA(\tau),
\label{vecQ1}
\\
&&{\bf q}(\tau,\tau') = \frac{e}{c}\frac{\int_{\tau'}^{\tau}{\bf
A}(\tau'')d\tau''}{\tau-\tau'}= \frac{e}{\omega}\frac{\vF(\tau)-\vF(\tau')}{\tau-\tau'}.\label{smq}
\end{eqnarray}
For the saddle point equation (\ref{saddle:gen}) we have the following explicit expression:
\begin{equation}
\label{R:saddle1} \vP^2(\tau-x_s)-\vQ^2(\tau-x_s,\tau)=0.
\end{equation}

One sees from Eq.~(\ref{Int1_xs}) for the terms $\cI_s[f_\vp^{(0)}]$ on the right-hand side
of Eq.~(\ref{intEq:f1}) that the oscillating (exponential) terms of the function $f_\vp^{(1)}(\tau)$
are partly determined through the phase functions $\varphi(\tau,x_s)$, which depend on the saddle
points $x_s$. Thus, the desired function $f_\vp^{(1)}(\tau)$ can be expressed as a sum,
\begin{equation}
f_\vp^{(1)}(\tau) = \sum_s g_{\vp,s}^{(1)}(\tau)e^{i\varphi(\tau,x_s)-iS_\vp(\tau)/\hbar},
\label{f1}
\end{equation}
where we have introduced a set of functions $g_{\vp,s}^{(1)}$ corresponding to each saddle point $x_s$.
Substitution of the form (\ref{f1}) for $f_\vp^{(1)}$ into Eq.~(\ref{intEq:f1}) gives the following equation for the pre-exponential functions $g_{\vp,s}^{(1)}$:
\begin{eqnarray}
\label{Eq:h}
&& \sum_{s} h_s(\tau)e^{i\varphi(\tau,x_s)}=0,\\
&& h_s(\tau) = \bigg(B_0[\cE_s^{(1)}(\tau)]-\frac{i}{\hbar}\sqrt{2m\cE_s^{(1)}(\tau)}\bigg)
g_{\vp,s}^{(1)}(\tau) \nonumber \\
&& \phantom{h_s(\tau)} - \frac{g_\vp^{(0)}(\tau-x_s)}
{\alpha_0\big[x_s^3\beta(\tau,x_s)\big]^{1/2}},
\label{h}
\end{eqnarray}
where the set of functions $\cE^{(1)}_s$ replaces $\cE^{(1)}$
in Eq.~(\ref{intEq:f1}). Comparison of the exponential factors in Eq.~(\ref{f1}) with that
in Eq.~(\ref{f:gen}) gives the following definition for $\cE^{(1)}_s(\tau)$:
\begin{eqnarray}
\cE_s^{(1)}(\tau) &=& \epsilon-\hbar\omega\frac{d}{d\tau}\big[\varphi(\tau,x_s)-S_\vp(\tau)/\hbar\big]
\nonumber \\
&=& \vQ^2(\tau,\tau-x_s)/(2m).\label{Es1}
\end{eqnarray}
In order to proceed, we assume, that any two different solutions, $x_s$ and $x_{s'}$, of
Eq.~(\ref{R:saddle1}) do not merge with variation of $\tau$ and, moreover, they are such that the
inequality,
\begin{equation} \label{ineq}
\Big|\frac{d}{d\tau}[\varphi(\tau,x_s)-\varphi(\tau,x_{s'})]\Big|\gtrsim u_p/(\hbar\omega),
\end{equation}
is fulfilled for the range of values of $\vp$ and parameters of the field $\vF(t)$ considered in this paper. [The validity of Eq.~(\ref{ineq}) can be justified by a numerical analysis of Eq.~(\ref{R:saddle1}) (cf.~Section~\ref{s.p.eq.}).] Under this assumption, the exponential factors in Eq.~(\ref{Eq:h}) can be considered as quasi orthogonal functions in the following sense:
\begin{equation*}
\Big|\int h_s(\tau)e^{i[\varphi(\tau,x_s)-\varphi(\tau,x_{s'})]}d\tau\Big|
\ll \Big|\int h_s(\tau)d\tau\Big|, \quad s\neq s'.
\end{equation*}
Therefore, without losing accuracy, we can consider only the trivial solution of Eq.~(\ref{Eq:h}),
$h_s(\tau)=0$, which from Eq.~\eqref{h} gives a set of uncoupled equations for the functions
$g_{\vp,s}^{(1)}(\tau)$.

Finally, taking into account Eqs.~(\ref{g0}) and~\eqref{Es1}, the preexponential factors
$g_{\vp,s}^{(1)}(\tau)$, that determine the first-order correction $f_\vp^{(1)}(\tau)$ in
Eq.~(\ref{f1}), can be expressed in terms of two field-free elastic scattering amplitudes
(\ref{ampl_el:s}) with different, field-dependent momenta:
\begin{equation}
\label{g1}
g_{\vp,s}^{(1)}(\tau) =\frac{\kappa\cA[P(\tau-x_s)]\cA[Q(\tau,\tau-x_s)]}
{\alpha_0\sqrt{x_s^3\beta(\tau,x_s)}}.
\end{equation}

The most remarkable consequences of Eqs.~(\ref{g0}) and (\ref{g1}) are that (i) both results
involve an exact (within effective range theory), non-Born field-free scattering amplitude ${\cal
A}(p)$ with laser-modified momentum; and (ii) the result (\ref{g1}) involves this amplitude twice.
Fact (ii) allows us to call the approximate result (\ref{f1}) ``the rescattering approximation.''
Thus the existence of laser-induced re-collisions in laser-assisted collision processes becomes
apparent already on the level of the QES wave function $\Phi_\vp(\vr,t)$, in which the
electron-atom dynamics and its modification by a strong laser field are completely described by the
function $f_\vp(t)$. The low-frequency analysis of the exact TDER equation (\ref{intEq:f}),
presented in this Section, allows us to obtain analytic closed-form expressions for the LAES
amplitude (\ref{amplitude:f}) corresponding to the zero-order [Eqs.~(\ref{f0}), (\ref{g0})] and
rescattering [Eqs.~(\ref{f1}), (\ref{g1})] approximations for $f_\vp(\tau)$ and, therefore, for the
scattering state $\Phi_\vp(\vr,t)$.

\section{The zero-order (Kroll-Watson) approximation for the LAES cross section}

Using the zero-order approximation $f_\vp(\tau)\approx f_\vp^{(0)}(\tau)$ [where $f_\vp^{(0)}(\tau)$ is given by Eqs.~(\ref{f0}) and (\ref{g0})],
we obtain for the LAES amplitude (\ref{amplitude:f}) the expression:
\begin{equation}
\label{ampl0_int}
\cA^{(0)}_n(\vp,\vp_n) = \frac{1}{2\pi}\int_0^{2\pi}\cA[P(\tau)]
e^{in\tau + i\Delta_n(\tau)}d\tau,
\end{equation}
where
\begin{eqnarray*}
&&\Delta_n(\tau)=[S_{\vp_n}(\tau)-S_{\vp}(\tau)]/\hbar= \rho\cos(\tau-\varphi_\vt),\\
&&\rho=\frac{|e|F}{m\hbar\omega^2}|\ve\cdot\vt|,\;\;
\varphi_\vt = \arg(\ve\cdot\vt),\;\; \vt=\vp_n-\vp,
\end{eqnarray*}
and the scalar product $(\ve\cdot\vt)$ is defined in accordance with Eq.~(\ref{scal-prod}). For the more general case of $l$-wave scattering, a low-frequency analysis of the TDER equations leads to the expression (\ref{ampl0_int}) for the scattering amplitude in which $\cA[P(\tau)]$ is replaced by $\cA^{(l)}[\vP(\tau),\vP_n(\tau)]$, where $\vP_n(\tau)=\vp_n - (e/c)\vA(\tau)$,
\begin{equation}
\label{ampl_el}
\mathcal{A}^{(l)}(\vp_i,\vp_f) = \frac{(2l+1)(k_ik_f)^{l}\mathcal{P}_l(\cos\theta_{if})}
{-1/a_l+r_lk_i^2/2 - ik_i^{2l+1} },
\end{equation}
$k_{i,f}=|\vp_{i,f}|/{\hbar}$, $\mathcal{P}_l(x)$ is a Legendre polynomial, and
$\theta_{if}=\angle(\vp_i,\vp_f)$. Later, we will omit the superscript $(l)$ denoting the
amplitude for field-free scattering (\ref{ampl_el}), $\cA(\vp_i,\vp_f)\equiv
\cA^{(l)}(\vp_i,\vp_f)$, bearing in mind that $\cA(\vp_i,\vp_f)$ contains information about the
spatial symmetry of the bound state supported by the scattering potential. Note that the amplitude
$\cA^{(s)}(p)$ for elastic $s$-wave scattering in Eq~(\ref{ampl_el:s}) is isotropic and depends only
on the modulus of the initial momentum. Thus, if necessary, the difference between $\cA^{(s)}(p)$
and $\cA^{(l\neq 0)}(\vp_i,\vp_f)$ will be indicated by using a different number of arguments.

It is important to note that the ``instantaneous'' amplitude $\cA[\vP(\tau),\vP_n(\tau)]$ that
replaces $\cA[P(\tau)]$ in Eq.~\eqref{ampl0_int} is not an elastic scattering amplitude (since
$|\vP(\tau)|\neq |\vP_n(\tau)|$). For the case of linear polarization ($\ell=1$)
Eq.~(\ref{ampl0_int}) corresponds to Eq.~(5.16) in Ref.~\cite{KW}, which involves the $T$-matrix
off the energy shell. For the case of elliptical polarization, results identical to
Eq.~(\ref{ampl0_int})  were obtained in Refs.~\cite{Milo96, Taulbjerg98}.

In the low-frequency limit ($\rho\gg 1$), the amplitude (\ref{ampl0_int}) can be evaluated analytically using uniform asymptotic approximation methods for integrals \cite{Bleistein, Wong} ~(cf.~Appendix~\ref{app2}):
\begin{equation}
\label{ampl0}
\cA^{(0)}_n=i^ne^{in\varphi_\vt}\bigg[ \cA_{+}J_n(\rho) + \cA_{-}\frac{i\rho J_n'(\rho)}{\sqrt{\rho^2-n^2}}\bigg],
\end{equation}
where $J_n'(x)$ is the derivative of the Bessel function $J_n(x)$,
\begin{equation}
\label{ampl0:Apm}
\cA_{\pm} = \frac{1}{2}\left[
\cA_{el}(\tau_{+}) \pm \cA_{el}(\tau_{-})
\right],
\end{equation}
and $\cA_{el}(\tau_{\pm})\equiv \cA[\vP(\tau_{\pm}),\vP_n(\tau_{\pm})]$, where $\tau=\tau_{\pm}$
are saddle points of the integrand in Eq.~(\ref{ampl0_int}) that satisfy the equation
\begin{equation}
\label{KW:sadEq}
\rho\sin(\tau-\varphi_\vt)=n.
\end{equation}
Because of the equality $|\vP(\tau_{\pm})|=|\vP_n(\tau_{\pm})|$, $\cA_{el}(\tau_{\pm})$ is the
on-shell amplitude for {\it elastic} field-free scattering with laser-modified momenta. This
modification serves to displace $\vp$ and $\vp_n$ by the shift $\Delta\vp_{\pm} =
(|e|/c)\vA(\tau_{\pm})$. For the classically allowed region $|n|\leq \rho$, Eq.~(\ref{KW:sadEq})
gives:
\begin{eqnarray}
\label{KW:saddles}
&&\tau_{\pm} = \varphi_\vt+\frac{\pi}{2}\pm\arccos\frac{n}{\rho}, \\
&&\Delta\vp_{\pm} = - \frac{m\hbar\omega}{2|\ve\cdot\vt|^2}
\Big[
\pm \xi[\vkap\times\vt]\sqrt{\rho^2-n^2} \nonumber \\
&&+n\big[ 2\ell(\veps\cdot\vt)\veps + (1-\ell)(\vt-(\vkap\cdot\vt)\vkap) \big]
\Big],
\label{KW:shifts}
\end{eqnarray}
where the degrees of linear ($\ell$) and circular ($\xi$) polarization are defined in Eq.~(\ref{ell-xi}). Note that for the case of critical geometry, when $(\ve\cdot\vt)\to 0$ (and thus $\rho\approx 0$), the result (\ref{ampl0}), based on a saddle point analysis of the integral (\ref{ampl0_int}), is not applicable.

The result (\ref{ampl0}) for $\cA^{(0)}_n$ and the corresponding cross section,
\begin{equation}
\frac{d\sigma_n^{(0)}(\vp,\vp_n)}{d\Omega_{\vp_n}}=\frac{p_n}{p}
\bigg| \cA_{+}J_n(\rho) + \cA_{-}\frac{i\rho J_n'(\rho)}{\sqrt{\rho^2-n^2}} \bigg|^2,
\label{cross0}
\end{equation}
may be simplified and reduced to the well-known Kroll-Watson formula~\cite{KW}
for the following particular cases of the laser polarization and the scattering geometry:

(i) For the case of linear polarization ($\ell=1$), we have $\Delta\vp_{+}=\Delta\vp_{-}=\Delta\vp$, where
\[
\Delta\vp=-m\hbar\omega n\, \frac{\veps}{(\veps\cdot\vt)},
\]
so that $\cA_{+}=\cA_{el}(\vp+\Delta\vp, \vp_n+\Delta\vp)$ and $\cA_{-}=0$, while
the cross section (\ref{cross0}) reduces to the original Kroll-Watson result~\cite{KW}:
\begin{equation}
\frac{d\sigma_n^{({\rm KW})}({\bf p},{\bf p}_n)}{d\Omega_{{\bf p}_n}}=\frac{p_n}{p} J^2_n(\rho)
\frac{d\sigma_{el}(\vP,\vP_n)}{d\Omega_{{\bf P}_n}},
\label{KW:cross}
\end{equation}
where $d\sigma_{el}/d\Omega = \big|\cA_{el}\big|^2$ is the exact cross section for field-free
elastic scattering and $\vP\equiv\vp+\Delta\vp$, $\vP_n\equiv\vp_n+\Delta\vp$. Note that the momentum shift $\Delta\vp$ for the case of
linear polarization  remains real in the classically forbidden region $|n|>\rho$.

(ii) For the cases of forward and backward scattering
along the major axis of the polarization ellipse ($\vp\|\vp_n\|\veps$), $\Delta\vp_{\pm}$
in Eq.~\eqref{KW:shifts} reduces as follows:
\begin{equation}
\Delta\vp_{\pm}  =  - \frac{m\hbar\omega}{|\vt|}
\Big[ \veps n 
\pm  \frac{\xi}{1+\ell}[\vkap\times\veps]\sqrt{\rho^2-n^2}\Big].
\label{KW:shifts(ii)}
\end{equation}
The collinearity of the vectors $\vp$, $\vp_n$, and $\vt$ gives the following
relations: $|\vP(\tau_{+})|=|\vP(\tau_{-})|$, $\vP(\tau_{\pm})=-\vP_n(\tau_{\mp})$.
Thus, $\cA_{-}=0$, $\cA_{+}=\cA_{el}(\tau_{+})=\cA_{el}(\tau_{-})$, and the LAES cross
section is given by Eq.~(\ref{KW:cross}) with $\vP = \vp + \Delta\vp_{\pm}$ and
$\vP_n = \vp_n + \Delta\vp_{\pm}$, where $\Delta\vp_{\pm}$ is given by Eq.~\eqref{KW:shifts(ii)}.
(This result is the same using either $\Delta\vp_{+}$ or $\Delta\vp_{-}$.)

(iii) For forward or backward scattering in the polarization plane for a circularly
polarized field ($\ell=0$), Eq.~\eqref{KW:shifts} gives
\begin{equation}
\Delta\vp_{\pm} = - \frac{m\hbar\omega}{|\vt|^2}
\Big[ \vt n \pm \xi[\vkap\times\vt]\sqrt{\rho^2-n^2}
\Big].
\label{KW:shifts(iii)}
\end{equation}
With $\Delta\vp_{\pm}$ given by Eq.~\eqref{KW:shifts(iii)}, the same analysis as for case (ii) is then valid.

\begin{figure}
\center
\includegraphics[width=\linewidth]{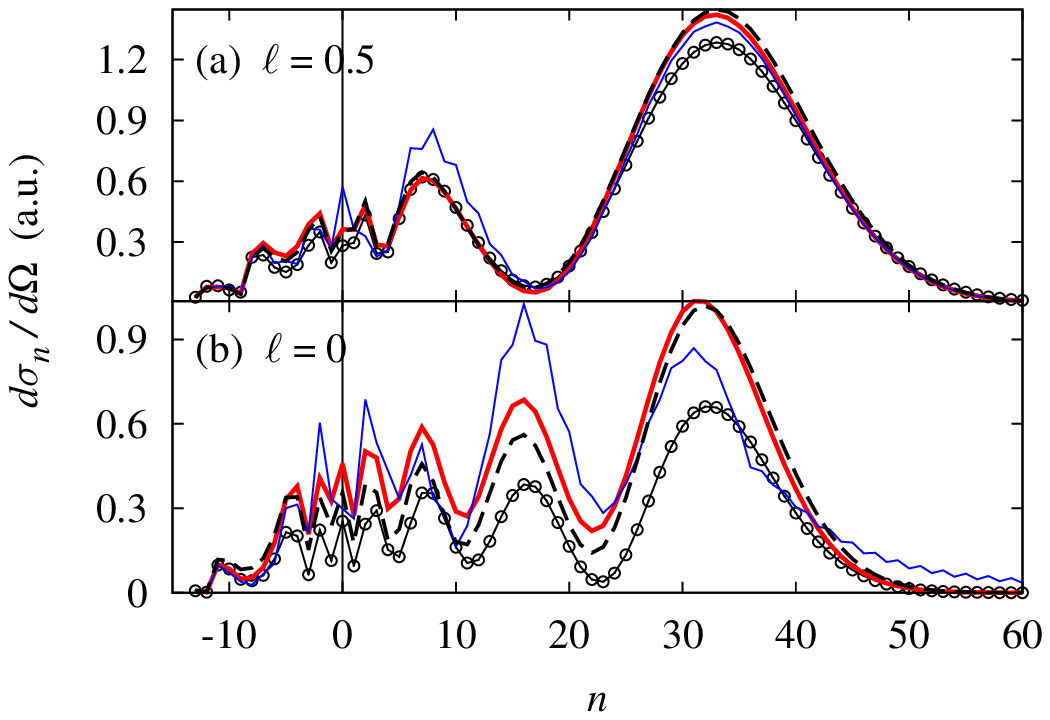}
\caption{(Color online)
Differential cross section $d\sigma_n({\bf p},{\bf p}_n)/d\Omega_{{\bf p}_n}$  for
laser-assisted $s$-wave $e$-H scattering in the polarization plane ($\vp\|\veps$,
$\vp_n\perp\vkap$) for a scattering angle $\theta\equiv\angle(\vp,\vp_n)=20^{\circ}$ in a
CO$_2$-laser field with $\hbar\omega=0.117$\,eV ($\lambda=10.6\,\mu$m) and intensity
$I=2.5\times10^{11}$\,W/cm$^2$. The incident electron energy is $E=1.58$\,eV and $n$ is the number
of photons absorbed ($n>0$) or emitted ($n<0$). Results are shown for two laser polarizations: (a)
Elliptical polarization, with $\eta=+0.58$ ($\ell=0.5$); (b) circular polarization, with $\eta=+1$
($\ell=0$). Circles: exact TDER results [cf.~Eqs.~\eqref{s:amplitude},~\eqref{cross}]; dashed
lines: results using the approximate amplitude~(\ref{ampl0_int}); thick solid (red) lines:
Eq.~\eqref{cross0}; thin solid (blue) lines: peaked impulse approximation (PIA) result of
Ref.~\cite{Taulbjerg98}. \label{fig1}}
\end{figure}

Note that other analytic expressions for the scattering amplitude (\ref{ampl0_int}) were obtained
in Refs.~\cite{Milo96,Taulbjerg98}. The LAES amplitude in the low-frequency approximation
introduced by Madsen and Taulbjerg \cite{Taulbjerg98} [labelled the ``peaked impulse
approximation'' (PIA)] has a form similar to Eq.~(\ref{ampl0}), but involves the Anger and Weber
functions~\cite{abramovitz} [cf.~Eq.~(\ref{A:Weber}) in Appendix~\ref{app2}]. In Fig.~\ref{fig1} we
compare the PIA result of Ref.~\cite{Taulbjerg98} with the analytic result (\ref{ampl0}), the
integral expression~(\ref{ampl0_int}) (within the effective range approximation),  and the exact
TDER results. The effective range theory parameters are those for $e$-H scattering:
$|E_0|=0.755$\,eV, $\kappa=0.236$\,a.u., $a_0=1.453\kappa^{-1}$, and $r_0=0.623\kappa^{-1}$. One
sees in Fig.~\ref{fig1} that the zero-order approximation (\ref{ampl0_int}) for the LAES amplitude
reproduces well the oscillation pattern in the LAES spectrum. It follows from Eq.~(\ref{ampl0})
that these oscillations are well approximated by the Bessel function and its derivative; they
originate from an interference of two classical electron trajectories corresponding to two
different times of collision, $\tau_+$ and $\tau_-$. In contrast, the result of
Ref.~\cite{Taulbjerg98} exhibits an additional sharp oscillatory structure for $d\sigma_n/d\Omega$
as a function of $n$ that stems from properties of the Weber function; they do not have any
physical meaning.

As may be seen from Eq.~(\ref{KW:saddles}), the two real saddle points $\tau_{\pm}$ coalesce at the
cutoff of the classically allowed region (i.e., for $n=\rho$). In the classically forbidden
region~($|n|>\rho$), the solutions of the saddle point equation~(\ref{KW:sadEq}) and the
corresponding momentum shifts~(\ref{KW:shifts}) become complex, so we analytically continue the
result (\ref{ampl0}) to this case. However, the complex displacements of momenta in the elastic
scattering amplitude may cause some non-physical features in the LAES cross section. Thus, for
example, for electron scattering with absorption or emission of $|n|>\rho\gg 1$ laser photons, the
condition $\vP^2(\tau_{\pm})/(2m)=-|E_0|$ may be satisfied for appropriate laser parameters and
geometry of the incident and outgoing electrons. For such conditions, the amplitude $\cA_{el}$ has
a pole, which corresponds to some point $\tau=\tau^{(p)}$  (or to more than one point) on the
complex plain of $\tau$. The coalescence of one of the saddle points $\tau_{\pm}$ with the point
$\tau^{(p)}$ leads to the appearance of a non-physical resonant-like enhancement of the LAES cross
section. (This fact is exhibited most clearly for the case of forward scattering and circular
polarization.) Thus, for the general case of elliptical polarization, the result (\ref{ampl0}) has
limited applicability in the classically forbidden region. For this case, an alternative analytic
result, suggested in Ref.~\cite{Taulbjerg98}, is obtained within an additional weak-field
approximation and, therefore, is not applicable for the description of strong laser field effects,
such as the plateau structures in LAES spectra.

\begin{figure}
\center
\includegraphics[width=\linewidth]{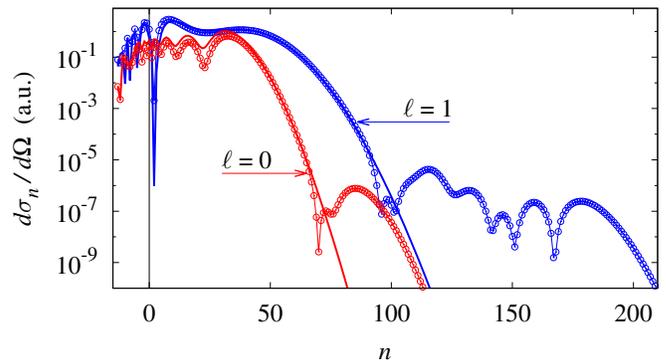}
\caption{(Color online)
The same as in Fig.~\ref{fig1}, but for the cases of linear ($\ell=1$) and circular
($\ell=0$) polarization, and for a larger range of $n>0$. Circles: exact TDER results
[cf.~Eqs.~\eqref{s:amplitude},~\eqref{cross}]; solid lines: results using the analytic
amplitude~(\ref{ampl0}). \label{fig1:log}}
\end{figure}

In Fig.~\ref{fig1:log} we present LAES spectra for $e$-H scattering in linearly and
circularly polarized CO$_2$-laser fields. The field intensity, electron energy, and scattering
geometry are the same as in Fig.~\ref{fig1}. For both of these two limiting cases of the laser
polarization, $\ell=1$ and $\ell=0$, as well as for the general case of elliptical polarization
($0\leqslant\ell \leqslant 1$), the zero-order (Kroll-Watson) approximation~(\ref{ampl0}) for the
LAES-amplitude does not describe the high-energy part of the LAES spectra (i.e., the rescattering
plateau), for which a proper account of laser-induced electron re-scattering from the potential
$U(r)$ is required~\cite{jetpl02,Milo04}. For the low-energy plateau, the result (\ref{ampl0}) is
in good agreement with the exact TDER results, except for the case of the critical geometry [for
which $\ve\cdot(\vp_n-\vp)=0$], as exhibited, e.g., by the pronounced suppression of the LAES cross
section within the Kroll-Watson approximation as compared to the exact result for $n=2$ and
$\ell=1$ (cf.~Fig.~\ref{fig1:log}). This discrepancy is due to the fact that the scattering angle
$\theta=20^\circ$ is close to the critical angle, $\theta_{cr}=21.05^\circ$, for the channel $n=2$.

\section{The rescattering approximation for the LAES amplitude}

The rescattering correction $\cA_n^{(1)}$ to the zero-order result~(\ref{ampl0}) for the LAES amplitude,
\begin{equation}
\label{Aresc}
\cA_n({\bf p},{\bf p}_n)\approx\cA_n^{(0)}+\cA_n^{(1)},
\end{equation}
follows upon substituting $f_\vp(\tau)=f_\vp^{(1)}(\tau)$
[where $f_\vp^{(1)}(\tau)$ is given by Eqs.~(\ref{f1}) and~\eqref{g1}] into Eq.~(\ref{amplitude:f}) to obtain:
\begin{eqnarray}
&&\cA_n^{(1)} = \sum_s\cA_{n,s}^{(1)}, \nonumber
\end{eqnarray}
\begin{eqnarray}
&&\cA_{n,s}^{(1)} = \frac{1}{2\pi\alpha_0} \int_0^{2\pi}\cA[P(\tau_s')] \cA[Q(\tau,\tau_s')]
\frac{e^{i\phi_s(\tau)}d\tau}{\sqrt{x_s^3\beta(\tau,x_s)}},\;\;\;\; \label{R:ampl_int}
\end{eqnarray}
where we have defined $\tau_s'\equiv \tau-x_s$ and, using Eq.~\eqref{Sp},
\begin{equation}
\label{R:phase}
\phi_s(\tau) \equiv \varphi(\tau,x_s) + n\tau - \frac{e}{m\hbar\omega^2}(\vp_n-\vp)\cdot\vF(\tau),
\end{equation}
where the functions $\varphi(\tau,x)$ and $\beta(\tau,x)$ are defined in Eqs.~(\ref{Int:phase}) and
(\ref{R:beta}) respectively, and $x_s=x_s(\tau)$ is defined  implicitly by Eq.~(\ref{R:saddle1}).

For the case of $l$-wave scattering, our analysis of the rescattering correction to the LAES
amplitude yields again an expression like~(\ref{R:ampl_int}), but with the field-free
$s$-wave scattering amplitude $\cA(p)$~[cf.~Eq.~(\ref{ampl_el:s})] in the integrand
of~(\ref{R:ampl_int}) replaced by $\cA(\vp_i,\vp_f)$~[cf.~Eq.~(\ref{ampl_el})]:
\begin{eqnarray*}
\cA[P(\tau_s')]\to\cA[\vP(\tau_s'),\vQ(\tau_s',\tau)],\\
\cA[Q(\tau,\tau_s')]\to \cA[\vQ(\tau,\tau_s'),\vP_n(\tau)],
\end{eqnarray*}
where $\vQ(\tau,\tau')$ is defined by Eqs.~\eqref{vecQ1} and~\eqref{smq}.

The dominant contributions to the integral (\ref{R:ampl_int}) come from the vicinity of the saddle points $\tau=\tau_k$, which satisfy the equation:
\begin{equation}
\label{R:saddle2}
2m\hbar\omega\frac{d\phi_s}{d\tau}\Big|_{\tau=\tau_k}=
\vP_n^2(\tau_k)-\vQ^2(\tau_k,\tau_k-x_s)=0.
\end{equation}
[In deriving Eq.~\eqref{R:saddle2}, use has been made of the relations $n\hbar\omega=(\vp_{n}^2-\vp^2)/(2m)$, $d\vF/d\tau=(\omega/c)\vA(\tau)$, and Eq.~\eqref{R:saddle1}.]
The saddle point equations (\ref{R:saddle1}) and (\ref{R:saddle2}) comprise a system of coupled
equations having a transparent physical meaning. Upon colliding with an atom at the time moment
$\tau_{s,k}' = \tau_k - x_s$, the electron changes its momentum $\vp$ to a field-dependent
``intermediate'' momentum $\vq(\tau_k,\tau_{s,k}')$, which ensures the condition for return of the
electron by the laser field back to the atom at the time moment $\tau_k$ followed by a rescattering.
The set of points $x_s$ determines the excursion times of the returning  electron along different
closed classical trajectories, while Eqs.~(\ref{R:saddle1}) and (\ref{R:saddle2}) represent the
energy conservation laws at the times of the first and second collisions. The argument $\vQ$ of
the field-free amplitude $\cA$ in Eq.~(\ref{R:ampl_int}) is the instantaneous kinetic
momentum of the electron in the laser field in the ``intermediate'' state with canonical momentum $\vq$
[cf.~Eq.~(\ref{vecQ1})].
\vspace{-0.3cm}

\subsection{Analysis of the saddle-point equations \label{s.p.eq.}}

To evaluate the integral in Eq.~(\ref{R:ampl_int}), it is instructive to
analyze first the solutions of the system of coupled saddle point equations (\ref{R:saddle1}) and (\ref{R:saddle2}).
Using dimensionless quantities, this system may be represented  as follows:
\begin{eqnarray}
\label{R:saddle11}
\vgamma^2 - \vnu^2 + 2(\vgamma-\vnu)
\cdot\Im\big(\ve\, e^{-i\tau'}\big)=0,\\
\label{R:saddle22}
\vgamma_n^2 - \vnu^2 + 2(\vgamma_n-\vnu)
\cdot\Im\big(\ve\, e^{-i\tau}\big)=0,
\end{eqnarray}
where $\vgamma\equiv\omega\vp/(|e|F)$, $\vgamma_n\equiv\omega\vp_n/(|e|F)$, and
$\vnu\equiv\vnu(\tau,\tau')=\omega\vq(\tau,\tau')/(|e|F)$.

Despite the fact that Eqs.~(\ref{R:saddle11}) and (\ref{R:saddle22})
are very similar, their solutions in the plane of variables $\tau$ and $\tau'$
(or, alternatively, $\tau$ and $x$, where $x=\tau-\tau'$) differ because
of the different ranges of the parameters $\gamma$ and $\gamma_n$.
Indeed, rescattering effects become important in the region of the LAES spectrum where
``direct'' scattering is classically forbidden, i.e., beyond the region of validity of the
Kroll-Watson result, where $\gamma_n>\sqrt{2(1+\ell)}-\gamma$
(for $\vgamma_n\|\vgamma\|\veps$)~\cite{circ05, cutoffs05}.
On the other hand, rescattering effects are most pronounced for low incident electron energy, i.e.,
$E\lesssim 2u_p$ or $\gamma\lesssim 1$.

\begin{figure}
\center
\includegraphics[height=0.68\linewidth]{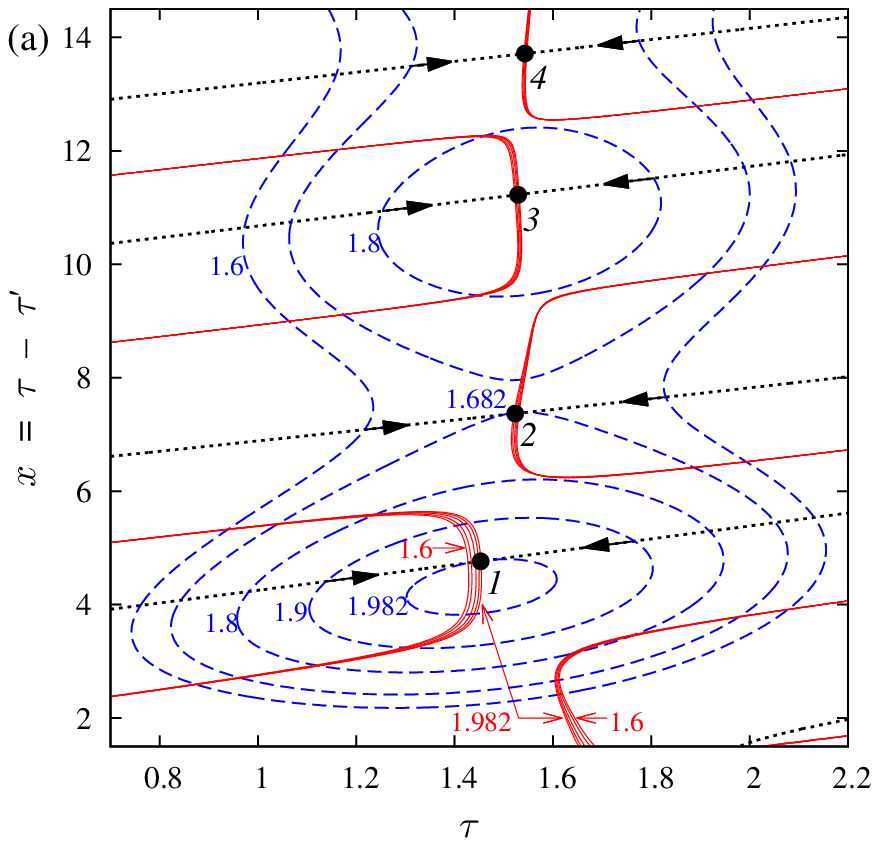}\hspace{-1mm}
\includegraphics[height=0.68\linewidth]{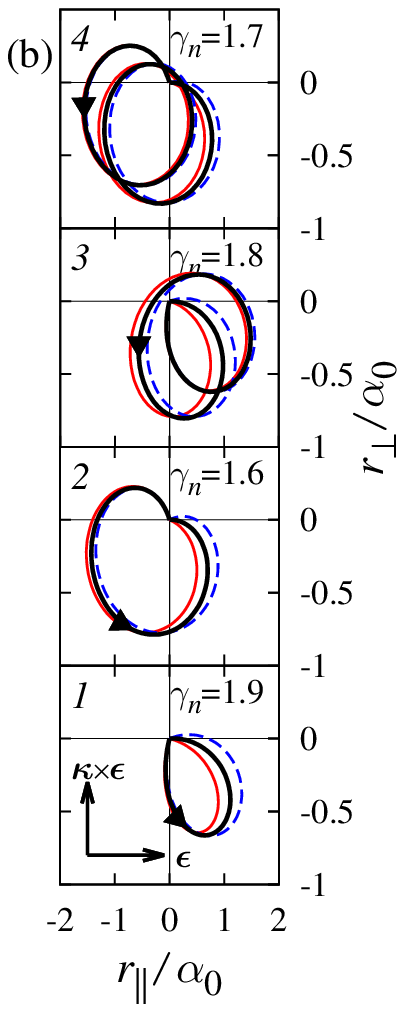}
\caption{(Color online)
(a) The solutions of Eqs.~(\ref{R:saddle11}) (dotted lines), (\ref{R:saddle22}) (dashed lines), and
(\ref{R:diff2}) (solid lines) for different values of $\gamma_n$, indicated in the figure near the
corresponding curve, $\gamma=0.6$, $\vgamma\|\vgamma_n\|\veps$, and polarization
$\eta=0.5$. The black arrows show the direction of movement of the coalescing solutions of the coupled
equations system (\ref{R:saddle11}), (\ref{R:saddle22}) with increasing $\gamma_n$. The corresponding
coalescence points [the solutions of the system of Eqs.~(\ref{R:saddle11}),~(\ref{R:diff2})] are indicated by the black dots labeled by the numerals $s=\textsl{1,2,3,4}$. (b) The classical closed trajectories of the electron in the polarization plane of the field $\vF(t)$. Thick solid (black) line: the coalesced (extremal) trajectories corresponding to the solutions $s=\textsl{1,2,3,4}$ of Eqs.~(\ref{R:saddle11}), (\ref{R:diff2}). Thin solid (red) and thin dashed (blue) lines: the short and long coalescing trajectories corresponding to the solutions of Eqs.~(\ref{R:saddle11}), (\ref{R:saddle22}).    \label{fig:saddle}}
\end{figure}

The numerical solutions of Eq.~(\ref{R:saddle11}) for $\gamma=0.6$ ($\vgamma\|\veps$) and
Eq.~(\ref{R:saddle22}) for different values of $\gamma_n$ ($\vgamma_n\|\veps$) for the case of
elliptical polarization with $\eta=0.5$ is shown in Fig.~\ref{fig:saddle}(a).
Fig.~\ref{fig:saddle}(a) illustrates the fact that, for the range of parameters considered,
Eq.~(\ref{R:saddle22}) has at most two real solutions $\tau^{(s)}_{\pm}$ on the trajectory
$x=x_s(\tau)$ of each solution of Eq.~(\ref{R:saddle11}). With increasing $\gamma_n$, the
points $\tau^{(s)}_{\pm}$ tend toward each other and coalesce at $\tau=\tau_s$ for
$\gamma_n=\gamma_{n}^{(s)}$. For example, the point \textsl{1} ($\tau_1=1.453, x_1=4.764$) in
Fig.~\ref{fig:saddle} corresponds to $\gamma_n=\gamma_{n}^{(1)}=1.982$, while the point \textsl{2} ($\tau_2=1.523, x_2=7.368$) corresponds to $\gamma_{n}^{(2)}=1.682$.

The coalescence of two real solutions, $\tau^{(s)}_{+}=\tau^{(s)}_{-}=\tau_s$, at
$\gamma_n=\gamma^{(s)}_{n}$ and their disappearance for $\gamma_n>\gamma^{(s)}_{n}$ means that the first derivative of $\phi_s(\tau)$ has a local minimum at $\tau=\tau_s$, while $\tau$ and $x$ vary along the trajectory of the solution $x=x_s(\tau)$. Thus the point $\tau=\tau_s$,
$x=x_s\equiv x_s(\tau_s)$ satisfies two coupled equations: Eq.~(\ref{R:saddle11}) and
$d^2\phi_s/d\tau^2=0$. The latter equation may be written as:
\begin{equation}
\label{R:diff2}
(\vnu - \vgamma_n)\cdot\Re\big(\ve\,e^{-i\tau}\big)(\tau-\tau') + \boldsymbol{\mathcal{Q}}^2 -
\boldsymbol{\mathcal{Q}}\cdot\boldsymbol{\mathcal{Q}}'\,\frac{d\tau'}{d\tau} = 0,
\end{equation}
where the following notations have been used:
\begin{eqnarray*}
&&\boldsymbol{\mathcal{Q}}=\vnu(\tau,\tau')+\Im\big(\ve\,e^{-i\tau}\big),\\ &&\boldsymbol{\mathcal{Q}}'=\vnu(\tau,\tau')+\Im\big(\ve\,e^{-i\tau'}\big),
\end{eqnarray*}
and where $d\tau'/d\tau$ is determined implicitly by Eq.~(\ref{R:saddle11}):
\begin{equation*}
\frac{d\tau'}{d\tau} = \frac{\boldsymbol{\mathcal{Q}}\cdot\boldsymbol{\mathcal{Q}}'}
{(\vgamma-\vnu)\cdot\Re\big(\ve\,e^{-i\tau'}\big)(\tau-\tau') + \boldsymbol{\mathcal{Q}}'^2}.
\end{equation*}
As one sees in Fig.~\ref{fig:saddle}(a), the solution $(\tau_s,x_s)$ of the system of Eqs.~(\ref{R:saddle11}) and~(\ref{R:diff2}) depends only weakly on $\gamma_n$.

The solutions $(\tau_s,x_s)$ may be grouped in pairs, labeled by two consecutive (odd and even)
integer subscripts $s$ [with the solutions $(\tau_s,x_s)$ enumerated in order of
increasing values of $x_s$, starting with $s=1$]. Analysis of the system of Eqs.~(\ref{R:saddle11}),~(\ref{R:diff2}) shows that the odd- and even-numbered solutions of each pair correspond respectively to greater and smaller values of $\gamma_n^{(s)}$. Moreover, the first pair of
solutions (i.e., $s=1,2$) provide two limiting values for $\gamma_n^{(s)}$: for
$\gamma_n>\gamma_n^{(1)}\equiv \gamma_{n,\max}$ the system (\ref{R:saddle11}), (\ref{R:saddle22})
does not have real solutions [the derivative $d\phi_s(\tau)/d\tau$ as a function of $\tau$ and $s$
has a global minimum at the point $(\tau_1,x_1)$], while the two saddle points $\tau^{(s)}_{\pm}$ do not coalesce for $\gamma_n<\gamma_n^{(2)}$. All other solutions $(\tau_s,x_s)$ correspond to
intermediate values of $\gamma_n^{(s)}$. A similar alternation of $\gamma_n^{(s)}$ with increasing $x_s$ exists also in the analysis of the ATI process and was described within the semiclassical rescattering model in Ref.~\cite{Becker02}.

Considering the classical motion of the electron in the laser field $\vF(t)$ described by
Newton's equation, $m\ddot{\vr}=-e\vF(t)$, a closed classical trajectory may be found for each
solution of the saddle point equations (\ref{R:saddle11}) and (\ref{R:saddle22}). For the geometry
$\vgamma_n\|\vgamma\|\veps$ and an elliptically polarized laser field, these trajectories lie in the
polarization plane ($\vr=r_{\|}\veps+r_{\bot}[\vkap\times\veps]$) and are shown in
Fig.~\ref{fig:saddle}(b) for different values $\gamma_n$. The two different rescattering times,
$\tau^{(s)}_{+}$ and $\tau^{(s)}_{-}$, correspond to the long and short trajectories respectively,
while the coalescence point $(\tau_s,x_s)$ corresponds to the extreme trajectory with
$\gamma_n=\gamma_n^{(s)}$. The smallest value of $x_s$~(i.e., $x_1$) is the return time of the
electron along the shortest extreme closed path. During its motion along this shortest trajectory,
the electron gains the maximal classical kinetic energy $E_{n,\max} = 2u_p\gamma_{n,\max}^{\,2}$.

With increasing $s$ (for $x_s\gg 1$), the solutions $\tau_s$ tend to a constant value (independent of $s$), while the sets of solutions $x_s$ with odd and even $s$ become equidistant:
$(x_{s+2}-x_s)\to2\pi$. This fact is easily verified by considering the solution of
Eqs.~(\ref{R:saddle11}) and (\ref{R:diff2}) in the limit $x=\tau-\tau'\gg~1$. For this case,
assuming $|\vgamma\cdot\ve|\neq 0$ and $|\vgamma_n\cdot\ve|\neq 0$, the system (\ref{R:saddle11}),
(\ref{R:diff2}) reduces to the much simpler system,
\begin{equation*}
\vgamma^2+2\vgamma\cdot\Im\big(\ve\,e^{-i(\tau-x)}\big)=0,\quad
\vgamma_n\cdot\Re\big(\ve\,e^{-i\tau}\big)=0,
\end{equation*}
which has the following solution:
\begin{eqnarray}
\label{R:tau:x>>1}
&&\tau = \varphi_{\vgamma_n}+\frac{\pi}{2},\\
&&x_{2k-(1\pm1)/2} = \varphi_{\vgamma_n}-\varphi_{\vgamma} + 2\pi k
\pm \arccos\frac{\vgamma^2}{2|\vgamma\cdot\ve|},\;\;\;
\label{R:x:x>>1}
\end{eqnarray}
where $\varphi_{\vgamma_n}=\arg(\vgamma_n\cdot\ve)$, $\varphi_{\vgamma}=\arg(\vgamma\cdot\ve)$. The
approximate results (\ref{R:tau:x>>1}), (\ref{R:x:x>>1}) are in reasonable agreement with the
numerical solutions of Eqs.~(\ref{R:saddle11}), (\ref{R:diff2}) beginning from the third pair of
points $(\tau_s,x_s)$ (for the example presented in Fig.~\ref{fig:saddle}, the relative error for
$\tau_{3}$ and $x_{3}$ is less than $3\%$ and $1\%$ respectively, while for $\tau_{4}$ and $x_{4}$ the error is less than $2\%$ and $0.6\%$).

Finally, we note that the solutions $(\tau_s,x_s)$ with even $s$ do not contribute to the high-energy region near the rescattering plateau cutoff, while they are important for the low-energy part of the rescattering plateau. The boundary energy, $\overline{E}_n$, between these two regions of the LAES spectrum is governed by the parameter $\overline{\gamma}_n$, which is the limiting value of $\gamma_n^{(s)}$ as $s\to\infty$, where $\gamma_n^{(2k-1)}$ for odd $s$ approaches $\overline{\gamma}_n$ from above, while $\gamma_n^{(2k)}$ for even $s$ approaches it from below. The equation for $\overline{\gamma}_n$ follows from Eq.~(\ref{R:saddle22}): $\overline{\gamma}_n^{\,2}=2|\overline{\vgamma}_n\cdot\ve|$. Using the parametrization (\ref{scal-prod}) for the scalar product $(\overline{\vgamma}_n\cdot\ve)$, the boundary energy $\overline{E}_n=2u_p\overline{\gamma}_n^{\,2}$ can be expressed as follows:
\begin{equation}
\overline{E}_n = 4u_p\sin\theta_{\vp_n}(1+\ell\cos2\phi_{\vp_n}),
\label{bound_energy}
\end{equation}
where $\theta_{\vp_n}$ and $\phi_{\vp_n}$ are the polar and azimuthal angles for the vector $\vp_n$ (or
$\vgamma_n$) in the basis $(\veps,[\vkap \times\veps],\vkap)$.

\subsection{Analytic formulas for the scattering amplitude}

Due to the coalescence of the two saddle points $\tau^{(s)}_{\pm}$ for each $s$, the ordinary saddle point method must be modified in order to  evaluate analytically the integral in Eq.~(\ref{R:ampl_int}) (which determines the LAES amplitude within the rescattering approximation). For this purpose we use the modification suggested in Ref.~\cite{NikRitus} and used recently to obtain factorized results for HHG~\cite{JPB2009} and ATI~\cite{analit_atd} yields. This modification consists in approximating the phase factor $\phi_s(\tau)$ by a cubic polynomial in the neighborhood of the point $\tau=\tau_s$, followed by removing from the integral (\ref{R:ampl_int}) the slowly-oscillating pre-exponential factor at $\tau=\tau_s$ and extending the range of integration to $\pm\infty$. The amplitude $\cA_n^{(1)}$ can then be evaluated analytically in terms of an Airy function, $\Ai(x)$~\cite{abramovitz}. The standard uniform approximation (in which one approximates the smooth pre-exponential factor by a linear function in the interval between the points $\tau=\tau^{(s)}_{\pm}$)~\cite{Bleistein,Wong}) gives approximately the same accuracy of results, but leads to cumbersome formulas, which are less suitable for further analyses and physical interpretations.

As discussed above, the function $\phi_s(\tau)$ is approximated as follows:
\begin{eqnarray}
\phi_s(\tau) &&\approx \tilde{\phi}_s + \frac{\vP_n^2(\tau_s) - \vQ^2(\tau_s,\tau_s')}{2m_e\hbar\omega}(\tau-\tau_s)
\nonumber \\
&&+ \frac{\alpha_s u_p}{3\hbar\omega}(\tau-\tau_s)^3,
\label{R:phase:ex}
\end{eqnarray}
where $\tau_s' =\tau_s - x_s$, $\tilde{\phi}_s\equiv\phi_s(\tau_s)$,  and the dimensionless factor
$\alpha_s$ is proportional to the third derivative of $\phi_s(\tau)$ at $\tau=\tau_s$, where in calculating this derivative one must take into account the $\tau$ dependence of $x_s(\tau)$, defined implicitly by Eq.~(\ref{R:saddle11}).  One obtains
\begin{equation}
\alpha_s = 2(\vnu_s-\vgamma_n)\cdot\Im(\ve\,e^{-i\tau_s}) + \Delta\alpha_s,
\label{R:alpha}
\end{equation}
where $\vnu_s\equiv\vnu(\tau_s,\tau_s')$ and
\begin{equation*}
\Delta\alpha_s = \frac{d^3}{d\tau^3}\Big[(x_s(\tau)-x_s)
(\vnu-\vnu_s)^2\Big]\Big|_{\tau=\tau_s}.
\end{equation*}
The explicit form of $\Delta\alpha_s$ is cumbersome. It is not presented here because numerical evaluation shows that it gives only a minor contribution to the final results.

Evaluating now the integral (\ref{R:ampl_int}), we take into account that the amplitudes
$\cA[\vP(\tau_s'),\vQ(\tau_s',\tau)]$ and $\cA[\vQ(\tau,\tau_s'),\vP_n(\tau)]$ depend only weakly on
$\tau$ in the neighborhood of the saddle points $\tau_{\pm}^{(s)}$
[which satisfy Eqs.~(\ref{R:saddle1}), (\ref{R:saddle2})]. Thus the amplitude $\cA$, evaluated at
$\tau=\tau_s$, can be replaced by the (on shell) amplitude $\cA_{el}$ of field-free elastic electron scattering. The result for the LAES amplitude $\cA_n^{(1)}$ is:
\begin{equation}
\label{ampl1}
\cA_n^{(1)}=\frac{1}{\alpha_0}\sum_{s=1}^{\infty}
D_s\cA_{el}(\vP^{(s)},\vQ'^{(s)}) \cA_{el}(\vQ^{(s)},\vP_{n}^{(s)}),
\end{equation}
where $\vP^{(s)}\equiv\vP(\tau_s')$, $\vP_n^{(s)}\equiv\vP_n(\tau_s)$,
$\vQ'^{(s)}\equiv\vQ(\tau_s',\tau_s)$, and $\vQ^{(s)}\equiv\vQ(\tau_s,\tau_s')$. The factors $D_s$ in
Eq.~(\ref{ampl1}) are expressed in terms of the Airy function:
\begin{equation}
\label{R:D}
D_s = \left(\frac{\hbar\omega}{u_p}\right)^{1/3}
\frac{e^{i\tilde{\phi}_s}\mathrm{Ai}(\zeta_s)}
{\alpha_s^{1/3}\sqrt{x_s^3\beta_s}},
\end{equation}
where $\beta_s\equiv\beta(\tau_s)$ is given by Eq.~(\ref{R:beta}), and
\begin{equation}
\label{R:arg_Ai}
\zeta_s = \frac{\big[(\vP_n^{(s)})^2 - (\vQ^{(s)})^2\big]/(2m)}
{u_p[\alpha_s(\hbar\omega/u_p)^2]^{1/3}}.
\end{equation}

The expression (\ref{ampl1}) may be simplified after further analysis and some additional
approximations. First, in accordance with the above analysis of the solutions of the saddle point equations, the sum over $s$ in Eq.~(\ref{ampl1}) can be split into separate sums over odd $s$ and even $s$. The sum over even $s$ contributes to the scattering amplitude only in the low-energy part of the rescattering plateau defined by $E_n<\overline{E}_n$ [cf.~Eq.~(\ref{bound_energy})]. Second, the contribution of each succeeding term of the sum in Eq.~(\ref{ampl1}) decreases because the coefficient $D_s$ decreases as $D_s\sim x_s^{-3/2}$. Furthermore, each succeeding odd ($s=2k+1$) term contributes negligibly to the scattering amplitude in the region $\gamma_n>\gamma_n^{(s)}$ because the Airy function $\Ai(\zeta_s)$ decreases exponentially for $\zeta_s>-1.019$. Thus we assume that the term with $s=1$ gives the dominant contribution in the region of rescattering plateau cutoff, that the term with $s=2$ contributes most to the region of the onset of the plateau, and that other terms (with higher $s$) give corrections in the intermediate region. Finally, the amplitude for field-free elastic scattering is a smooth function of its arguments and changes only slightly with respect to variations of $s$ having the same parity, owing to the quasi-equidistant feature of the solutions $(\tau_s,x_s)$ [cf.~Eqs.~(\ref{R:tau:x>>1}) and (\ref{R:x:x>>1})]. These considerations allow us to approximate the amplitude $\cA_n^{(1)}$ by separating the summation over $s$ in Eq.~(\ref{ampl1}) into two sums (over odd and even $s$) and by removing the slowly varying amplitudes $\cA_{el}$, evaluated at the proper momenta, from under each summation. Since the main contributions to the sum (\ref{ampl1}) are given by the first terms of the two separate summations (for odd and even $s$), we assume that the momenta are the corresponding instantaneous kinetic momenta, evaluated at the (dimensionless) times $(\tau_1,\tau_1')$ for the odd $s$ sum: $\vP=\vP^{(1)}$, $\vP_n=\vP_n^{(1)}$, $\vQ'=\vQ'^{(1)}$, $\vQ=\vQ^{(1)}$, and evaluated at the times $(\tau_2,\tau_2')$ for the even $s$ sum: $\tilde{\vP}=\vP^{(2)}$, $\tilde{\vP}_n=\vP_n^{(2)}$, $\tilde{\vQ}'=\vQ'^{(2)}$,$\tilde{\vQ}=\vQ^{(2)}$. The result is:
\begin{eqnarray}
\nonumber
\cA_n^{(1)}(\vp,\vp_n) & = & \frac{1}{\alpha_0}
\big[ D^{(o)}\cA_{el}(\vP,\vQ') \cA_{el}(\vQ,\vP_{n})
\\
& + & D^{(e)}\cA_{el}(\tilde{\vP},\tilde{\vQ}') \cA_{el}(\tilde{\vQ},\tilde{\vP}_{n})
\big],
\label{ampl1:2terms}
\end{eqnarray}
where $D^{(o)}=\sum_{k=0}^{\infty}D_{2k+1}$, $D^{(e)}=\sum_{k=1}^{\infty}D_{2k}$.

The approximate result (\ref{ampl1:2terms}) [as well as the more accurate result (\ref{ampl1})] shows that the LAES amplitude with account of rescattering effects is given by a sum of factorized terms: all effects of the scattering potential $U(r)$ are collected in the two exact amplitudes $\cA_{el}$ for field-free elastic electron scattering, while the factors $D_s$ [defined by Eq.~(\ref{R:D}) in terms of an Airy function] depend only on the laser parameters. Therefore, neither the scattering amplitude nor the LAES cross section can be factorized over the entire rescattering plateau region as a product of only two (laser and atomic) factors; however, such a factorization becomes possible in the high-energy part of the rescattering plateau, due to the negligible contribution of the second term in Eq.~(\ref{ampl1:2terms}) in this region.


\section{Factorization of the LAES cross section in the rescattering plateau region}

\subsection{Three-step formula for the LAES cross section }

In the high-energy part of LAES spectrum, we can neglect the second term of
Eq.~(\ref{ampl1:2terms}) for the LAES amplitude in the rescattering approximation as well as the
first (Kroll-Watson) term in Eq.~(\ref{Aresc}). Substituting Eq.~(\ref{ampl1:2terms}) into
Eq.~(\ref{cross}), we obtain a factorized result for the LAES differential cross section in the
high-energy region of the rescattering plateau:
\begin{equation}
\label{R:CS}
\frac{d\sigma_n^{(r)}(\vp,\vp_n)}{d\Omega_{\vp_n}} =
\frac{d\sigma_{el}(\vP,\vQ')}{d\Omega_{\vQ'}}
\mathcal{W}(\vp,\vp_n)
\frac{d\sigma_{el}(\vQ,\vP_{n})}{d\Omega_{\vP_{n}}},
\end{equation}
where the factor $\mathcal{W}(\vp,\vp_n)$,
\begin{equation}
\label{R:W}
\mathcal{W}(\vp,\vp_n) = \frac{p_n}{\alpha_0^2\,p}\Big|\sum_{k=0}^{\infty}D_{2k+1}\Big|^2,
\end{equation}
depends on the momenta $\vp$ and $\vp_n$ of the incident and scattered electrons through the
explicit dependence of the instantaneous momentum $\vP_{n}^{(s)}$ [$=\vp_n- e\vA(\tau'_s)/c$] in
the argument of the Airy function in Eq.~(\ref{R:D}), and through the implicit dependence of the
times $\tau_s=\tau_s(\vp,\vp_n)$ and $\tau_s'=\tau_s'(\vp,\vp_n)$ on the momenta $\vp$ and $\vp_n$.
Since Eq.~(\ref{R:CS}) was obtained as a simplified, low-frequency version of the exact quantum
results for the scattering problem, its expression in terms of three factors provides a convincing
quantum justification  of the classical three-step rescattering scenario of the LAES process for
the general case of an elliptically polarized laser field.

The cross section $d\sigma_{el}(\vP,\vQ')/d\Omega_{\vQ'}$ in Eq.~(\ref{R:CS}) describes the elastic
scattering of an electron with initial momentum ${\vp}$ from the potential $U(r)$ at the time
moment $t'=\tau_1'/\omega$. Since the collision takes place in the presence of a field ${\bf
F}(t)$, this term involves (instead of the momentum $\vp$) the laser-modified instantaneous
momentum $\vP$ of the electron at the moment of collision. The scattering direction is given by the
vector $\vQ'$, which is determined only by the vector potential of the elliptically polarized laser
field and has the sense of an intermediate ``kinetic momentum'' of the electron in an
``intermediate'' state, immediately after the elastic scattering event at the moment $t'$. From
this state the electron  starts to move in the laser field up to the moment  of the second
scattering (or rescattering).  The cross section $d\sigma_{el}(\vP,\vQ')/d\Omega_{\vQ'}$, involving
the instantaneous momenta $\vP$ and $\vQ'$, describes elastic scattering (since $|\vQ'|=|\vP|$),
while the initial momentum $\vp$ changes to $\vq$ ($|\vp|\neq |\vq|$). In order to ensure the
condition for return of the electron back to the origin [where the magnitude of the potential
$U(r)$ is maximal] at the moment $t$, the vector $\vq=\vq(\tau_1,\tau_1')$ depends on two times:
the time $t'$ of the first collision and the time $t=\tau_1/\omega$ of rescattering. The result of
the rescattering at the moment $t$ is that the electron with the intermediate momentum $\vq$
rescatters along the direction of the final (detected) momentum $\vp_n$. This event is described in
Eq.~(\ref{R:CS}) by the cross section $d\sigma_{el}(\vQ,\vP_{n})/d\Omega_{\vP_{n}}$ for field-free
elastic scattering with instantaneous momenta $\vQ$ and $\vP_{n}$ (where $|\vP_{n}|=|\vQ|$).

The key factor in the factorized cross section (\ref{R:CS}) is the propagation factor ${\cal W}(\vp,\vp_n)$. This factor describes the motion of a free electron in the field ${\bf F}(t)$ for the time $\Delta t= t-t'$ resulting in the change of its initial kinetic momentum $\vP$ to $\vP_{n}$. Indeed, as is seen from the explicit form for $D_{s=2k+1}$ in Eq.~(\ref{R:D}), the expression (\ref{R:W}) for ${\cal W}(\vp,\vp_n)$ does not involve any dependence on the potential $U(r)$ and is determined completely by the free electron motion in the field ${\bf F}(t)$. Our numerical analysis shows that the sum over $k$ in Eq.~(\ref{R:W}) converges rapidly for arbitrary electron energy $E_n$ in the rescattering plateau region, so that only the first few terms in this sum over the saddle points contribute significantly. These terms effectively take into account both short and long closed trajectories of the electron in the laser field. These trajectories correspond to the two solutions, $\tau_{\pm}^{(s)}$, of the saddle point equations (\ref{R:saddle11}), (\ref{R:saddle22}) whose interference causes the oscillatory features in the LAES spectra, which originate mathematically from the behavior of the Airy function $\Ai(\zeta_s)$. The times $t_s=\tau_s/\omega$ and $\Delta t_s =x_s/\omega$, which govern the magnitude of $D_{s}$ in Eq.~(\ref{R:D}), are respectively the moment of rescattering and the excursion time for electron propagation along the closed trajectory corresponding to the extreme path for which the $s$th pair of short and long trajectories coalesce [as shown in Fig.~\ref{fig:saddle}(b)]. The numerator of the Airy function argument $\zeta_s$ in Eq.~(\ref{R:arg_Ai}) represents the difference between the kinetic energy of the electron with final momentum ${\bf p}_n$ and the maximum classical energy, $(\vQ^{(s)})^2/(2m)$, that can be gained by an electron with initial momentum ${\bf p}$ in the laser field before the rescattering event.

The physical interpretation of Eq.~(\ref{R:CS}) is most clear if we limit ourselves to the case of the high-energy plateau cutoff region in the LAES spectrum, for which only the first term of the sum in Eq.~(\ref{R:W}) dominates and the factor $\mathcal{W}$ involves only a single term, $D_1$:
\begin{equation}
\label{R:W1}
\mathcal{W}(\vp,\vp_n) \approx \frac{p_n}{\alpha_0^2\,p}|D_{1}|^2.
\end{equation}
For the case of linear polarization $(\ell =1)$, the factorization (\ref{R:CS}) with
$\mathcal{W}(\vp,\vp_n)$ given by Eq.~(\ref{R:W1}) coincides with that obtained in
Ref.~\cite{analit_laes}. The result (\ref{R:W1}) takes into account only the return of the electron
along the first pair of short and long closed classical trajectories in Fig.~\ref{fig:saddle}(b),
while the terms with $k\geq 1$ in the sum over $k$ in Eq.~(\ref{R:W}) determine the correction to
the propagation factor in Eq.~(\ref{R:W1}) due to electron returns along other ``odd'' (with
$s=2k+1$) pairs of short and long trajectories [cf.~Fig.~\ref{fig:saddle}(b)].

\subsection{Comparison with the exact TDER results}


\begin{figure}
\center
\includegraphics[width=\linewidth]{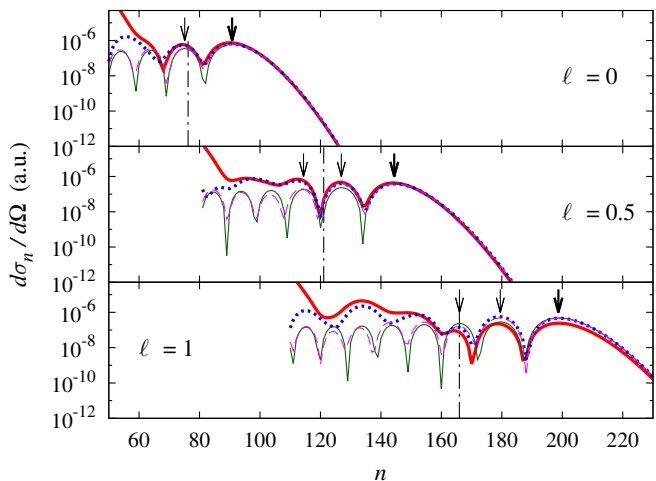}
\caption{(Color online)
LAES differential cross section for forward $s$-wave $e$-H scattering
($\vp\|\vp_n\|\veps$) as a function of the number $n$ of absorbed laser photons in an
elliptically polarized CO$_2$-laser field with $\hbar\omega=0.117$\,eV ($\lambda=10.6\,\mu$m) and
intensity $I=2.5\times10^{11}$\,W/cm$^2$ for three different degrees of linear polarization $\ell$
($= 0$, $0.5$, $1$), and incident electron energy $E=1.58$\,eV. Thick solid lines: exact TDER
results; dotted lines: result (\ref{ampl1}) for the LAES amplitude; dashed lines: the three-step
formula (\ref{R:CS}); thin solid lines: Eq.~(\ref{R:CS}) with approximation (\ref{R:W1}) for the
propagation factor. Vertical dotted-dashed lines mark the position of the boundary
[cf.~Eq.~(\ref{bound_energy})] between the two regions of the rescattering plateau. Arrows indicate
the positions of the interference maxima and the plateau cutoffs. \label{fig:ell}}
\end{figure}

In Figs.~\ref{fig:ell} and~\ref{fig:theta} we compare exact TDER results for $s$-wave scattering
(cf. Section~1 of Appendix~\ref{app1}) with the low-frequency analytic results (for effective range
theory parameters $a_0$ and $r_0$ corresponding to the case of $e$-H scattering). One sees that the
analytic result (\ref{ampl1}) for the scattering amplitude describes well the entire rescattering
plateau region of the LAES spectra [we find that the simpler two-term result (\ref{ampl1:2terms})
for $\cA_n^{(1)}(\vp,\vp_n)$ provides the same accuracy in describing the rescattering
plateau]. For the high-energy part of the plateau ($E_n>\overline{E}_n$), the three-step formula
(\ref{R:CS}) is in good agreement with the exact results. Moreover, the main contribution is given
by the term corresponding to the shortest excursion time of the electron along the closed
trajectory [cf.~Eq.~(\ref{R:W1})]. The account of the longer trajectories [given by the terms in
Eq.~(\ref{R:W}) with $k>0$] provides a correction to the result (\ref{R:W1}) in the spectral region
between $\overline{E}_n$ and the energy corresponding to the last (closest to the plateau cutoff)
oscillatory minimum.

Our analysis shows that the agreement between the analytic formula and the exact results in the
cutoff region worsens for $\ell\to 1$ (cf.~Fig.~\ref{fig:ell}). This fact is connected with the
loss of the contributions to the scattering amplitude of the intermediate QES-channels with
negative quasienergies $\epsilon_n = E+u_p+n\hbar\omega$ [cf.~Eq.~(\ref{s:iterative})] when the
saddle-point approximation for the exact TDER equations was made. The effect of the closed channels
on the LAES amplitude is not considered in this paper. We just note that the contributions of the
closed channels to the LAES cross section in the high-energy plateau region noticeably depends on
the laser intensity and the incident electron energy $E$ for a linearly polarized field and
disappears for the case of the circular polarization.

\begin{figure}
\center
\includegraphics[width=\linewidth]{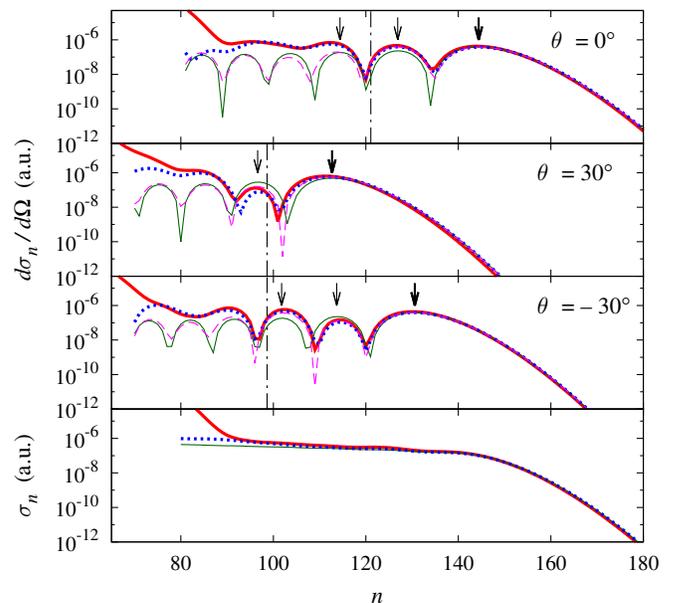}
\caption{(Color online)
The same as in Fig.~\ref{fig:ell}, but for scattering in the polarization plane
($\vp\|\veps$, $\vp_n\bot\vkap$) with $\ell=0.5$ and three different scattering angles
$\theta=\angle(\vp,\vp_n)$ ($ 0^\circ, \pm30^\circ$). Bottom panel: the LAES angle-integrated
differential cross section over the ``forward scattering'' hemisphere (see text).
\label{fig:theta}}
\end{figure}

The comparison of our analytic results with exact TDER results for $p$-wave scattering (cf.
Section~2 of Appendix~\ref{app1}) is presented in Figs.~\ref{fig:ell:p} and~\ref{fig:theta:p},
where the effective range theory parameters are those for $e$-F scattering: $|E_0|=3.401$\,eV,
$\kappa=0.500$\,a.u., $a_1=0.827\kappa^{-3},$ and $r_1=-4.417\kappa$. The intensity,
$I=6.92\times10^{12}$\,W/cm$^2$, of a mid-infrared laser field with $\hbar\omega=0.354$\,eV
($\lambda=3.5\,\mu$m) and the incident electron energy, $E=4.78$\,eV, are chosen so that the
ratios $u_p/(\hbar\omega)$ and $E/(\hbar\omega)$ have the same values as for $s$-wave scattering in
Figs.~\ref{fig:ell} and~\ref{fig:theta}. One sees that the accuracy of the analytic result
(\ref{ampl1}) for the scattering amplitude and of the three-step formula (\ref{R:CS}) for the LAES cross
section for $p$-wave scattering is as good as for the case of $s$-wave scattering (cf. Figs.~\ref{fig:ell}
and~\ref{fig:theta}).

\begin{figure}
\center
\includegraphics[width=\linewidth]{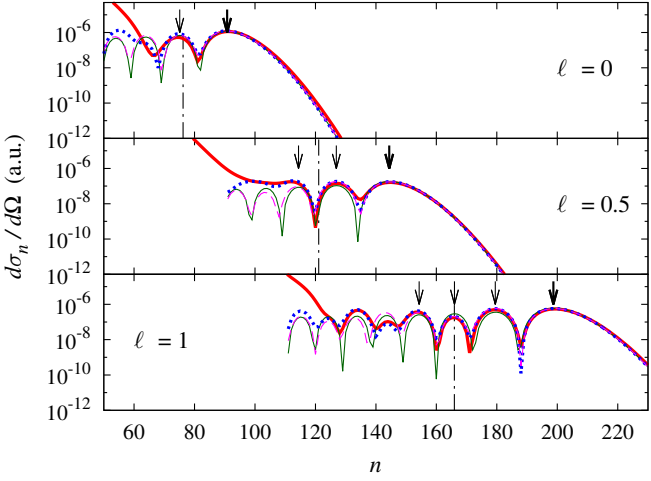}
\caption{(Color online)
The same as in Fig.~\ref{fig:ell}, but for $p$-wave $e$-F scattering in a laser field with
$\hbar\omega=0.354$\,eV ($\lambda=3.5\,\mu$m), $I=6.92\times10^{12}$\,W/cm$^2$, and incident
electron energy $E=4.78$\,eV. \label{fig:ell:p}}
\end{figure}

\begin{figure}
\center
\includegraphics[width=\linewidth]{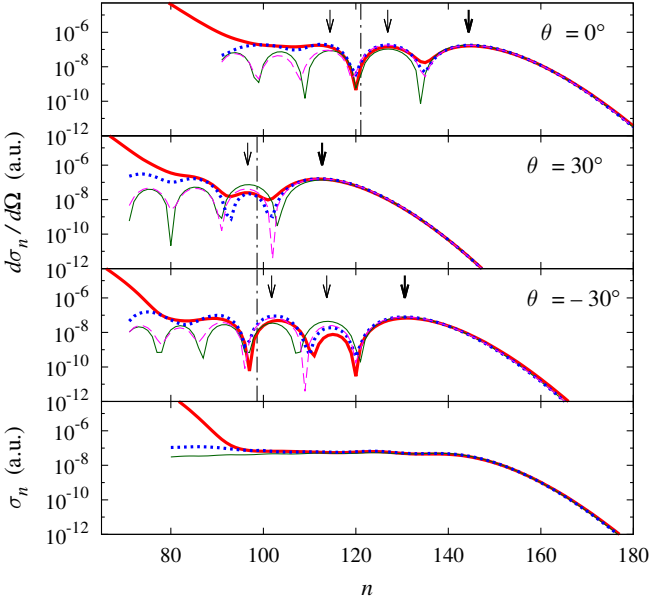}
\caption{(Color online)
The same as in Fig.~\ref{fig:theta}, but for $p$-wave $e$-F scattering. The laser field
parameters $\omega$ and $I$, and the electron energy $E$ are the same as in Fig.~\ref{fig:ell:p}.
\label{fig:theta:p}}
\end{figure}

\subsection{Discussion}

\begin{figure}
\center
\includegraphics[width=0.9\linewidth]{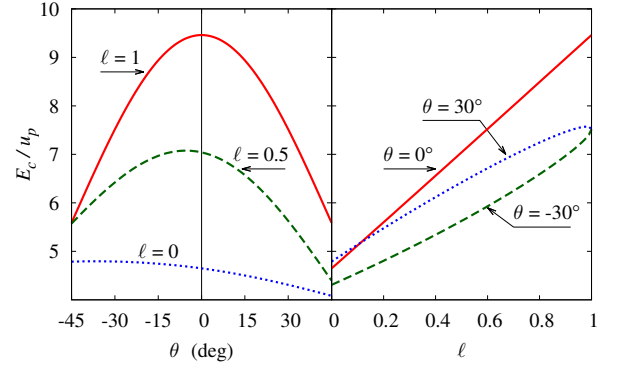}
\caption{(Color online)
The cutoff energy $E_c$ vs. scattering angle $\theta$ for different values of the linear
polarization degree $\ell$ (left panel) and $E_c$ vs $\ell$ for different $\theta$ (right panel).
The scattering geometry is $\vp\|\veps$, $\vp_n\bot\vkap$ and the laser parameters and the incident
electron energy are the same as in Fig.~\ref{fig:ell}. \label{fig:cutoff}} \vspace{-0.3cm}
\end{figure}

The analytic results (\ref{R:CS}) -- (\ref{R:W1}) allow one to explain all features of LAES spectra
in the region of the rescattering plateau, as shown in Figs.~\ref{fig:ell} and~\ref{fig:theta} for
$s$-wave ($e$-H) scattering and in Figs.~\ref{fig:ell:p} and~\ref{fig:theta:p} for $p$-wave ($e$-F)
scattering. Moreover, in the case that the field-free cross sections ${d\sigma_{el}}/{d\Omega}$
have a smooth energy dependence, these features are governed by the propagation factor
$\mathcal{W}(\vp,\vp_n)$ and are insensitive to the details of the potential $U(r)$. In particular,
the position of the plateau cutoff, as well as the positions of the maxima and minima in the
oscillation pattern below the plateau cutoff, are described quantitatively with high accuracy by
the properties of the Airy function $\Ai(\zeta_1)$ [where $\zeta_1$ is defined in
Eq.~(\ref{R:arg_Ai})] (cf.~similar analyses for high-energy HHG and ATI spectra in
Refs.~\cite{JPB2009,analit_atd}). If the energy difference in the numerator of $\zeta_1$ in
Eq.~(\ref{R:arg_Ai}) is positive, the Airy function (and hence the LAES cross section) decreases
exponentially with increasing $p_n$. In contrast, $\Ai(\zeta_1)$ oscillates for $\zeta_1<0$ with
the position of its first maximum at $\zeta_1 \equiv z_1=-1.019.$ This value of $\zeta_1$ thus
determines the position of the plateau cutoff ($E_{n,\max}\equiv E_c=2u_p\gamma_c^2$) in the LAES
spectrum, where $\gamma_c$ satisfies the transcendental equation obtained by equating $\zeta_1$ to
$z_1$:
\begin{eqnarray}
\nonumber
&&\vgamma_c^2-\vnu_1^2+2(\vgamma_c-\vnu_1)\cdot\Im(\ve\,e^{-i\tau_1}) \\
&&= z_1\alpha_1^{1/3}\left({\hbar\omega}/{u_p}\right)^{2/3},
\label{cutoff:Eq}
\end{eqnarray}
where $\alpha_1$ is given by Eq.~(\ref{R:alpha}), $\vnu_1\equiv\vnu(\tau_1,\tau_1')$, and
$(\tau_1,\tau_1')$ is the first solution (corresponding to the shortest return time $x_1$) of the
system of equations (\ref{R:saddle11}), (\ref{R:diff2}) with $\vgamma_n =\vgamma_c$. In other
words, the cutoff parameter $\vgamma_c$ is given by the joint solution of the coupled system of
Eqs.~(\ref{R:saddle11}), (\ref{R:diff2}), and (\ref{cutoff:Eq}). For an arbitrary ellipticity
(including the case of circular polarization), an analytic expression for $\vgamma_c$ may be found
only as a polynomial interpolation of the exact numerical solution of Eqs.~(\ref{R:saddle11}),
(\ref{R:diff2}), and (\ref{cutoff:Eq}) and, in general, this interpolation has a cumbersome form
because of its dependence on the many parameters of the problem (such as, e.g.,  the scattering
geometry, the scattering angle, the ellipticity, the incident electron energy, and the laser
intensity). Thus we show in Fig.~\ref{fig:cutoff} the numerical solutions of the transcendental
equations for $E_c$ for scattering in the polarization plane for different values of the
ellipticity and the scattering angle.

As shown in Fig.~\ref{fig:cutoff}, the cutoff position depends strongly on the scattering angle
$\theta$ for $\ell=1$: $E_c(\theta)\approx E_c(0)-7.9u_p\theta^2$ (cf.~Ref.~\cite{analit_laes}),
while the angular dependence of $E_c(\theta)$ becomes smoother with decreasing linear polarization
degree $\ell$. For forward scattering along the direction of the major axis of the polarization
ellipse, the dependence of $E_c(\ell)$ on $\ell$ is close to linear over a wide interval of
incident electron energies $E$ and laser intensities $I$ [$I=c F^2/(8\pi)$]: $E_c(\ell)/u_p \approx
a_1(E,F)+a_2(E,F)\ell$, where $a_{1,2}(E,F)$ are smooth functions of $E$ and $F$
(cf.~Fig.~\ref{fig:cutoff}).

Another noticeable effect seen in Fig.~\ref{fig:cutoff} is an asymmetry in the cutoff position with
respect to the sign of the angle $\theta$ for $\ell < 1$ (cf.~also Figs.~\ref{fig:theta}
and~\ref{fig:theta:p}). (For the geometry $\vp\|\veps$, $\vp_n\bot\vkap$, one has
$p_n\cos\theta=\vp_n\cdot\veps$ and $p_n\sin\theta=\vp_n\cdot[\vkap\times\veps]$, so that the
positive direction of $\theta$ coincides with the direction of the field rotation for $\eta>0$.)
This dichroic effect for the cutoff of the rescattering plateau in LAES spectra was predicted
in~Ref.~\cite{circ05}.

The oscillation pattern in the dependence of $\mathcal{W}(\vp,\vp_n)$ on $\vp_n$ originates from
the interference of two classical electron trajectories, which merge at the cutoff with the
shortest extremal trajectory and which were taken into account in evaluating the LAES
amplitude (cf.~discussion in Sec.~\ref{s.p.eq.}). This interference explains the oscillatory patterns
in the LAES spectra below the plateau cutoff (for $\zeta_1<z_1$), which are known from numerical
calculations (cf.~Ref.~\cite{circ05} and Figs.~\ref{fig:ell}~-- \ref{fig:theta:p}) and were
discussed in Refs.~\cite{analit_laes} and~\cite{Milo06}. [In Ref.~\cite{Milo06} the origin of the
oscillatory patterns as a consequence of the interference between real electron trajectories
was established by taking into account the scattering potential $U(r)$ perturbatively within the
strong-field and uniform approximations.]

The positions of the minima/maxima of the interference oscillations may be found in the same way as
for the cutoff position, i.e., by solving the system (\ref{R:saddle11}), (\ref{R:diff2}), and
Eq.~(\ref{cutoff:Eq}) for $\vgamma_n=\vgamma_{n,\min/\max}$, replacing $z_1$ by $z_k$ [where $z_k$
are the positions of the zeros and the maxima of $\Ai^2(\zeta_1)$]. For $k\geq 2$, the values of
$z_k$ are well approximated by equating to $\pi k/2$ the argument of the sine function in the
asymptotic form of $\Ai(-|\zeta_1|)$ for large $|\zeta_1|$~\cite{abramovitz},
\[
\Ai(-|\zeta_1|)\propto |\zeta_1|^{-1/4}\sin\left( \frac{2}{3}|\zeta_1|^{3/2}+\frac{\pi}{4} \right).
\]
The maxima/zeros of $\Ai^2(\zeta_1)$ (and hence the maxima/minima of $d\sigma_n/d\Omega$) correspond to odd/even $k$ in the relation
\[
\zeta_1=z_k=-0.25[2\pi(2k-1)]^{2/3},\quad k\geq 2.
\]
The estimated positions of a few maxima in the LAES spectra closest to the cutoff are indicated in
Figs.~\ref{fig:ell}~-- \ref{fig:theta:p} by arrows. One sees that these positions coincide well
with the positions of the maxima in the exact TDER results.

\begin{figure}
\center
\includegraphics[width=0.9\linewidth]{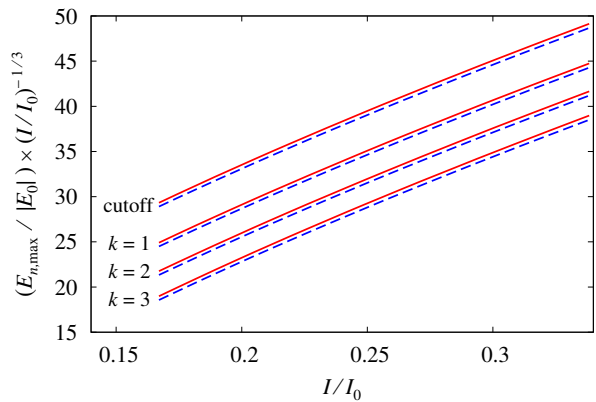}
\caption{(Color online)
Scaled positions ($E_{n,\max}$) of the plateau cutoff and of the oscillatory maxima
closest to the cutoff (marked by $k=1,2,3$) vs. scaled laser intensity $I/I_0$ for forward
scattering in a circularly polarized field and for two electron energies: $E=13.5\hbar\omega$
(solid lines) and $E=19.5\hbar\omega$ (dashed lines). [Scaled units of intensity,
$I_0=1.5\times10^{12}$\,W/cm$^2$, and of energy, $|E_0|=0.755$\,eV, correspond to $e$-H scattering
(see text). For the case of a CO$_2$-laser: $E=1.58$\,eV and 2.28\,eV.] \label{fig:cutoffI}}
\vspace{-0.5cm}
\end{figure}

We have found that the positions of the maxima or minima in the oscillatory LAES spectra depend on
the scattering angle and on the laser polarization in much the same way as shown for the cutoff
position, $E_c(\theta,\ell)$, in Fig.~\ref{fig:cutoff}.  However, the distance between the
positions of the maxima or minima for fixed $\ell$ and $\theta$ depends essentially only on the
laser intensity and scales as $I^{1/3}$. This fact is shown in Fig.~\ref{fig:cutoffI} for the case
of forward $e$-H scattering in a circularly polarized ($\ell=0$) field for two values of the
electron energy $E$. [The scaled unit of intensity, $I_0$, in Fig.~\ref{fig:cutoffI} is defined as
$I_0 =cF_0^2/(8\pi)$, where $F_0=\sqrt{8m|E_0|^3}/(|e|\hbar)$. Thus for $e$-H scattering
$(|E_0|=0.755$\,eV), $I_0=1.5\times10^{12}$\,W/cm$^2.$] Note that for a linearly polarized field,
the same intensity dependence for the positions of the maxima and minima was found analytically for
LAES~\cite{analit_laes} and for ATD~\cite{analit_atd} processes.

Because of the sensitivity of the oscillatory patterns in the LAES spectra to the scattering angle
(cf.~Figs.~\ref{fig:theta} and~\ref{fig:theta:p}), the angle-integrated spectra are smooth, as
shown in the bottom panels in Figs.~\ref{fig:theta} and~\ref{fig:theta:p}, in which the integration
was performed over the ``forward scattering'' hemisphere: $0\leq\theta_{\vp_n}\leq 180^\circ$,
$-90^\circ\leq\phi_{\vp_n}\leq 90^\circ$, where $\theta_{\vp_n}$ and $\phi_{\vp_n}$ are the polar
and azimuthal angles for the vector $\vp_n$. For this case, one sees in Figs.~\ref{fig:theta}
and~\ref{fig:theta:p} that the simple analytic result (\ref{R:CS}) with propagation factor
(\ref{R:W1}) provides good agreement with the exact TDER results over the entire rescattering
plateau.

\section{Conclusions and perspectives}

Nowadays the manifestation of field-free atomic dynamics in strong field processes and the
retrieval of information on this dynamics from the measured outcomes of laser-atom interactions are
attracting increasing interest. For HHG and ATI processes, this dynamical information can be
obtained theoretically most convincingly using well-developed algorithms for direct numerical
solution of the time-dependent Schr\"{o}dinger equation. However, for laser-assisted collisions,
numerical algorithms for calculating the scattering state wave function in an intense,
low-frequency laser field have not yet been developed, even for the case of linear laser
polarization. Moreover, the widely-used strong field approximation is not applicable for this
purpose since for an electron in the continuum it treats the scattering potential perturbatively,
using the Born approximation. Thus for collision problems, non-perturbative approximate theories or
exactly-solvable models play an essential role in providing a deeper understanding of the influence
of the scattering potential on laser-assisted collision processes.

In this paper, we have obtained quantum-mechanically (in the low-frequency limit) analytic
expressions for cross sections of electron scattering from a potential in the presence of an
elliptically polarized laser field using TDER theory, which permits one to obtain not only an exact
numerical solution for the LAES problem but also simple analytic results for a number of limiting
cases. Our analytic derivations are based on the analytic representation of the exact TDER
scattering state $\Phi_\vp(\vr,t)$ in Eq.~(\ref{scatt.state}) as a sum of two terms: the
``zero-order'' term, which corresponds to the low-frequency, Kroll-Watson result for the scattering
state [cf. Eq.~(5.12) in Ref.~\cite{KW}], and the ``rescattering correction,'' which takes into
account the strong laser field modifications of the electron interaction with the scattering
potential $U(r)$ beyond the Kroll-Watson approximation. Since the Kroll-Watson term in the LAES
cross section decreases exponentially beyond the classically-allowed region (for high $n$), the
rescattering correction becomes dominant there and describes perfectly the rescattering plateau in
the high-energy region of the LAES spectrum. The high accuracy of our analytic approximations for
the exact TDER LAES amplitude is demonstrated by comparison of analytic and exact numerical TDER
results for the ellipticity and angular dependences of LAES spectra for two different cases: $s$-wave scattering (corresponding to electron scattering from hydrogen or an alkali atom;
cf.~Figs.~\ref{fig:ell}, \ref{fig:theta} for $e$-H scattering) and $p$-wave scattering (corresponding to a halogen atom target; cf.~Figs.~\ref{fig:ell:p}, \ref{fig:theta:p} for $e$-F scattering).

The key results of this paper are the expression (\ref{ampl1}) for the LAES amplitude in the
rescattering approximation and the three-step formula (\ref{R:CS}) for the LAES cross section. The
factorized result (\ref{R:CS}) describes well the high-energy part of the rescattering plateau,
while the non-factorized LAES amplitude (\ref{ampl1}) [as well as the two-term result
(\ref{ampl1:2terms})] describes the LAES spectrum over the entire rescattering plateau region (cf.
Figs.~\ref{fig:ell}~-- \ref{fig:theta:p}). After substituting Eq.~(\ref{R:W1}) for the propagation
factor, the formula (\ref{R:CS}) provides a generalization of the result for a linearly polarized
laser field~\cite{analit_laes} to the case of nonzero driving laser ellipticity.

The major limitation of the TDER theory model is that it takes into account only a single
partial-wave scattering phase (in a given $l$-wave channel) for the potential
$U(r)$~\cite{comm,two-stateTDER}, whereas the entire set of phase shifts should be taken into
account in describing elastic electron scattering by a neutral atom. However, this deficiency is
compensated by the very clear and physically transparent interpretation of our key results
(\ref{ampl1}) and (\ref{R:CS}). Indeed, (i) the quantum-mechanically derived factorized formula
(\ref{R:CS}) agrees completely with the semiclassical three-step rescattering scenario for the LAES
process giving, in fact, a quantum ``replica'' (or quantum justification) of this scenario; (ii)
the account of rescattering effects in our analysis was performed non-perturbatively in the
potential $U(r)$, so that the results (\ref{ampl1}) and (\ref{R:CS}) contain the exact (non-Born)
amplitude and cross section for elastic electron scattering by the potential $U(r)$ within the
effective range theory; and (iii) the factors $D_s$ [cf.~Eq.~(\ref{R:D})] in Eq.~(\ref{ampl1}), as
well as the propagation factor $\mathcal{W}(\vp, \vp_n)$, do not involve any parameters of the
potential $U(r)$ and thus are valid for any atomic target. [In particular, our results for the
$s$-wave and the $p$-wave scattering show that these factors do not depend on the spatial symmetry
of a bound state (if it exists) in an atomic potential $U(r)$.] Therefore, it is reasonable to
expect that a generalization of Eqs.~(\ref{ampl1}) and~(\ref{R:CS}) beyond the TDER theory may be
performed quite straightforwardly, i.e., replacing the field-free scattering amplitudes $\cA_{el}$
in Eq.~(\ref{ampl1}) and the TDER cross sections $d\sigma_{el}/d\Omega$ in Eq.~(\ref{R:CS}) by the
amplitudes and cross sections for elastic electron scattering by a particular real atom obtained
from either experimental measurements or accurate theoretical calculations. Similar generalizations
of factorized TDER results for HHG~\cite{FMSERSPRL09} and ATI~\cite{analit_atd} yields to the case
of real atomic targets have been shown to provide fine agreement with results of accurate numerical
solutions of the time-dependent Schr\"{o}dinger equation for the plateau cutoff region in HHG and
ATI spectra. For LAES, the aforementioned generalization allows one to extend the formulas
(\ref{ampl1}) and (\ref{R:CS}) to the case of atomic targets (such as inert gases) which do not
support a bound state of an attached electron (i.e., a negative ion) in spite of the fact that the
description of LAES within the TDER theory presented in this paper is not applicable for such
cases. The use of the results (\ref{ampl1}) and (\ref{R:CS}) for such cases that go beyond the
present TDER theory will be described in a separate publication.

The results in this paper become inapplicable for resonant electron energies,
$E\approx\mu\hbar\omega-|E_0|-u_p$, at which the electron may be temporarily captured in a bound
state $\psi_{\kappa l m_l}(\vr)$ of the potential $U(r)$ by emitting $\mu$
photons~\cite{prl_laes09}, and for threshold energies, $E=k\hbar\omega,$ $k=1,2,\ldots$, at which
the LAES spectrum may be affected considerably by threshold phenomena, corresponding to the closing
(or opening) of the channel for stimulated emission of $k$ laser photons by the incident
electron~\cite{jetpl08}. Since both resonant and threshold phenomena have a purely quantum origin,
when the discreteness of the photon energy $n\hbar\omega$ is essential, these phenomena disappear
in the low-frequency approximation ($\hbar\omega\to 0$) used in the present work. An analysis of
resonant and threshold phenomena for the LAES process in an elliptically polarized laser field will
be published elsewhere.

Finally, we note that, even for the simplest geometry, $\vp\|\veps$, the ellipticity $\eta$ of the
laser field affects significantly the angular distribution (AD) of scattered electrons as compared
to the case of linear polarization, because it destroys the axial symmetry of the AD that exists
for $\eta=0$ with respect to the direction of $\veps$. In particular, the ADs for $\eta \neq 0$
differ substantially for $\eta=\pm|\eta|$, thus exhibiting an elliptic dichroism effect whose
detailed study for both the low-energy and the rescattering regions of the LAES spectrum is now in
progress.

\section*{ACKNOWLEDGMENTS}

This work was supported in part by RFBR Grant 
No.\,13-02-00420,  by NSF Grant No.\,PHY-1208059, and by the Russian Federation Ministry of
Education and Science (Contract No.\,14.B37.21.1937).


\appendix
\section{The matrix form of the TDER equations for the Fourier coefficients $f^{(lm_l)}_k(\vp)$ and the LAES amplitude \label{app1}}

\subsection{Results for $s$-wave scattering $(l=0)$}

Equation~(\ref{intEq:f}) can be converted into a system of inhomogeneous linear algebraic
equations for the Fourier coefficients $f_k(\vp)$ of the function $f_\vp(\tau)=\sum_k f_k(\vp) e^{-ik\tau}$:
\begin{equation}
\label{s:system} \sum_{s'}\mathcal{M}_{s,s'}(\epsilon+\delta\hbar\omega)f_{2s'+\delta}(\vp) = \kappa
c_{2s+\delta}(\vp), 
\end{equation}
where the symbol $\delta$ is equal to $0$ $(1)$ for an even (odd) $k$.
The inhomogeneous term in the system (\ref{s:system}) is
expressed in terms of Fourier coefficients of the wave function $\chi_\vp(\vr=0,\tau)$
[cf.~Eq.~(\ref{Volkov})]:
\begin{equation}
c_k(\vp) = i^k\mathcal{J}_{-k}^*\left(\frac{|e|F}{m\hbar\omega^2}(\ve\cdot\vp), \frac{\ell
u_p}{2\hbar\omega}\right), \label{s:ck}
\end{equation}
where $\mathcal{J}_n(z,x)$ is a generalized Bessel function:
\[
\mathcal{J}_n(z,x) = \sum_{p=-\infty}^\infty e^{i(n+2p)\arg(z)} J_{n+2p}(|z|)J_{p}(x).
\]
Therefore, the system
(\ref{s:system}) is equivalent to two separate (uncoupled) systems for
even and odd Fourier coefficients of the QES wave function $\Phi_\vp(\vr,t)$ at $r\to 0$.

The matrix elements $\mathcal{M}_{s,s'}(\epsilon)$ in Eq.~(\ref{s:system}) have the following form:
\begin{eqnarray} \label{Mssprime}
&& \mathcal{M}_{s,s'}(\epsilon) =
\cA^{-1}(\tilde{p}_{\,2s})\delta_{s,s'} - M_{s,s'}(\epsilon),
 \\
&& \cA(\tilde{p}_{\,2s}) = \frac{1}{- a_0^{-1} + r_0 k_{2s}^2/2 - ik_{2s} },\quad k_{2s}=\frac{\tilde{p}_{\,2s}}{\hbar},\hspace{3mm}
\label{s:ampl}\\
&& M_{s,s'}(\epsilon) = i^{s-s'}\sqrt{\frac{m\omega}{2\pi i\hbar}}
\int_0^{\infty}\frac{d\tau}{\tau^{3/2}} e^{i\epsilon_{s+s'}\tau/(\hbar\omega)}
\nonumber \\
&& \qquad \times \big[ e^{-i\lambda(\tau)} J_{s-s'}(\ell z(\tau))
-\delta_{s,s'} \big], \label{s:matrix}\\
&& \lambda(\tau) = \frac{u_p}{\hbar\omega}\Big(\tau-\frac{4}{\tau}\sin^2\frac{\tau}{2}\Big),
\nonumber \\
&& z(\tau) = \frac{u_p}{\hbar\omega}\Big(\sin\tau-\frac{4}{\tau}\sin^2\frac{\tau}{2}\Big),
\nonumber
\end{eqnarray}
where $J_n(x)$ is a Bessel function, and
the following notations are used in Eqs.~(\ref{Mssprime}) -- (\ref{s:matrix}): $\epsilon_n \equiv \epsilon + n\hbar\omega = E+u_p+n\hbar\omega$, $\tilde{p}_{n}=\sqrt{2m\epsilon_n}$. Note that only diagonal matrix elements $\mathcal{M}_{s,s'}$ contain the information on atomic dynamics [i.e., the field-free elastic scattering amplitude $\cA(\tilde{p}_{\,2s})$ for a ``momentum'' $\tilde{p}_{\,2s}$, which is imaginary for closed channels, with $\epsilon_{2s}<0$], while the non-diagonal elements ($s\neq s'$) depend only on the incident electron energy $E$ and the laser parameters.

In terms of the coefficients $f_k(\vp)$, the LAES amplitude (\ref{amplitude:f}) can be represented in
an alternative form~\cite{jetpl08}:
\begin{equation}
\label{s:amplitude} \mathcal{A}_n(\vp,\vp_n) = \kappa^{-1}\sum_{k=-\infty}^\infty
f_k(\vp)c_{k-n}^*(\vp_n).
\end{equation}

The low-frequency iterative solution of the integro-differential equation (\ref{intEq:f}),
presented in Section~\ref{low-frequency}, corresponds to the iterative account of the integral
terms $M_{s,s'}$ in Eq.~(\ref{s:matrix}) for solving the system (\ref{s:system}).
In the lowest order in $M_{s,s'}$, the solution of Eq.~(\ref{s:system}) is:
\begin{equation}
f_k(\vp)\approx \kappa\cA(\tilde{p}_k)\big[c_k - \sum_{s'}\cA(\tilde{p}_{k+2s'})M_{0,s'}(\epsilon_{k})
c_{k+2s'}\big], \label{s:iterative}
\end{equation}
The first term in the approximation (\ref{s:iterative}) corresponds to the zero-order approximation
(\ref{f0}) for the function $f_\vp(\tau)$, while the second term describes the rescattering
correction (\ref{f1}). However, we emphasize that the approximation (\ref{s:iterative}) is more
accurate than the low-frequency expansion (\ref{f01}) because the LAES amplitude
(\ref{s:amplitude})  [as well as the sum over $s'$ in Eq.~(\ref{s:iterative})] involves a summation
over all intermediate channels, including closed channels. Nevertheless, using the approximation
(\ref{s:iterative}) we are not able to provide a closed-form analytic expression for the LAES
amplitude. Finally, we note that all non-diagonal matrix elements $M_{s,s'}$ (with $s\neq s'$) are
equal to zero for a circularly polarized ($\ell=0$) field $\vF(t)$. In this case the sum over $s'$
in Eq.~(\ref{s:iterative}) contains only the single term with $s'=0$.

\subsection{Results for $p$-wave scattering $(l=1)$}

For $l=1$, matching the QES wave function (\ref{scatt.state}) [with $\Phi_\vp^{(sc)}(\vr,t)$ given
by Eq.~\eqref{wf:out:l}] to the small-$r$ boundary condition \eqref{BC:qes} results in the system
of three (for $\mu=0$, $\pm1$) coupled integro-differential equations for functions
$f^{(1\mu)}_{\vp}(\tau) = \sum_k f_k^{(1\mu)}(\vp) e^{-ik\tau}$ [cf.~Eq.~(\ref{intEq:f}) for the
case $l=0$]. This system can be converted into the following three matrix equations for the Fourier coefficients
$f_k^{(\mu)} \equiv f_k^{(1\mu)}(\vp)$:
\begin{eqnarray}
\label{p:eq1}
&& \sum_{s'}M^{(0)}_{s,s'}(\epsilon_\delta)f^{(0)}_{2s'+\delta} = \kappa^2c^{(0)}_{2s+\delta},\\
\label{p:eq2}
&& \sum_{s'}\!\left(
\begin{array}{cc}
\hspace{-1.5mm}\bar{M}^{(-1)}_{s,s'}(\epsilon_\delta)\hspace{-1mm} & \hat{M}^{(-1)}_{s,s'}(\epsilon_\delta) \hspace{-2mm}\\
\hspace{-1.5mm}\hat{M}^{(1)}_{s,s'}(\epsilon_\delta)\hspace{-1mm} & \bar{M}^{(1)}_{s,s'}(\epsilon_\delta)\hspace{-2mm}
\end{array}
\right)\hspace{-1mm}
\left(
\begin{array}{cc}
\hspace{-1mm}f^{(-1)}_{2s'+\delta}\hspace{-1mm} \\
\hspace{-1mm}f^{(1)}_{2s'+\delta}\hspace{-1mm}
\end{array}
\right)\! =
\kappa^2\!\left(
\begin{array}{cc}
\hspace{-1mm}c^{(-1)}_{2s+\delta}\hspace{-1mm} \\
\hspace{-1mm}c^{(1)}_{2s+\delta}\hspace{-1mm}
\end{array}
\right)\hspace{-1mm},\hspace{5mm}
\end{eqnarray}
where $\epsilon_\delta = \epsilon + \delta\hbar\omega$ and $\delta$ is equal to $0$ $(1)$ for an
even (odd) $k$, similarly to the result for $s$-wave scattering in Eq.~\eqref{s:system}. The
coefficients $c^{(\mu)}_k$ on the right-hand side of Eqs.~\eqref{p:eq1}, \eqref{p:eq2} can be
expressed in terms of the coefficients $c_k(\vp)$, given by Eq.~\eqref{s:ck}:
\begin{eqnarray*}
c^{(\mu)}_k(\vp) = \frac{p}{\hbar}\sqrt{4\pi}Y_{1\mu}^{*}(\hat{\vp})
c_k(\vp) + i\mu\sqrt{3(1+\ell)}\frac{|e|F}{4\hbar\omega} \nonumber \\
\times\Big[
\Big(\!1+\frac{\mu\xi}{1+\ell}\Big)c_{k-1}(\vp) - \Big(\!1-\frac{\mu\xi}{1+\ell}\Big)c_{k+1}(\vp)
\Big],
\label{p:ck}
\end{eqnarray*}
where the spherical harmonic $Y_{1\mu}(\hat{\vp})$ is defined as in Ref.~\cite{Varshalovich}.

The matrix elements $M_{s,s'}^{(0)}(\epsilon)$, $\bar{M}^{(\mu)}_{s,s'}(\epsilon)$ and $\hat{M}^{(\mu)}_{s,s'}(\epsilon)$ ($\mu=\pm1$)  in Eqs.~(\ref{p:eq1}) and~(\ref{p:eq2}) have the following form (cf. Ref.~\cite{TDER2008}):
\begin{eqnarray}
&& M^{(0)}_{s,s'}(\epsilon) = \Big(-\frac{1}{a_1} + \frac{r_1k_{2s}^2}{2} - ik_{2s}^3\Big) \delta_{s,s'} \nonumber \\
&&\qquad + \mathcal{C}\int_0^{\infty}\frac{d\tau}{\tau^{5/2}}
e^{i\epsilon_{s+s'}\tau/(\hbar\omega)}
\nonumber \\
&&\qquad \times
\big[e^{-i\lambda(\tau)}J_{s-s'}(\ell z(\tau)) -\delta_{s,s'} \big],
\label{p:matrix1} \\
&& \bar{M}^{(\mu)}_{s,s'}(\epsilon) = M^{(0)}_{s,s'}(\epsilon) +
\mathcal{C}
\int_0^{\infty}\frac{d\tau}{\tau^{3/2}} e^{i\epsilon_{s+s'}\tau/(\hbar\omega)-i\lambda(\tau)}
\nonumber \\
&&\qquad \times \big\{ [i\rho_1(\tau)+\mu\xi z(\tau)]J_{s-s'}(\ell z(\tau)) \nonumber \\
&&\qquad - \ell \rho_2(\tau)J_{s-s'}'(\ell z(\tau)) \big\},
\label{p:matrix2} \\
&& \hat{M}^{(\mu)}_{s,s'}(\epsilon) =
\mathcal{C}
\int_0^{\infty}\frac{d\tau}{\tau^{3/2}} e^{i\epsilon_{s+s'}\tau/(\hbar\omega)-i\lambda(\tau)}
\nonumber \\
&&\qquad \times \Big\{ -i\ell\rho_1(\tau) J_{s-s'}(\ell z(\tau))
+ \rho_2(\tau)\Big[J_{s-s'}'(\ell z(\tau)) \nonumber \\
&&\qquad  + \frac{\mu\xi(s-s')}{\ell z(\tau)}J_{s-s'}(\ell z(\tau))\Big] \Big\},
\label{p:matrix3}
\end{eqnarray}
where $J_n'(z)$ is the derivative of the Bessel function and the following notations are used:
\begin{eqnarray*}
&&\mathcal{C}=\frac{3i^{s-s'+1}}{\sqrt{2\pi i}}\Big(\frac{m\omega}{\hbar}\Big)^{3/2},\\
&&\rho_1(\tau)=\frac{u_p}{\hbar\omega}\left(\frac{4}{\tau^2}\sin^2\frac{\tau}{2} - \frac{2}{\tau}\sin\tau + \cos\tau\right),\\
&&\rho_2(\tau)=\frac{u_p}{\hbar\omega}\left(\frac{4}{\tau^2}\sin^2\frac{\tau}{2} - \frac{2}{\tau}\sin\tau + 1\right).
\end{eqnarray*}

Once the Fourier coefficients $f^{(\mu)}_k(\vp)$ are known, the exact TDER result for the $p$-wave LAES
amplitude is given by:
\begin{equation}
\label{p:amplitude} \mathcal{A}_n^{(l=1)}(\vp,\vp_n) = \kappa^{-2}\sum_{\mu=-1}^{1} \sum_{k=-\infty}^{\infty} f^{(\mu)}_k(\vp)c^{(\mu)*}_{k-n}(\vp_n).
\end{equation}

\section{The uniform asymptotic approximation of the integral~(\ref{ampl0_int}) \label{app2}}

In this Appendix, we describe the approach for the uniform asymptotic expansion of the integral~(\ref{ampl0_int}).
We note first that after replacing the integration variable $\tau$ in Eq.~(\ref{ampl0_int})
by $x=\tau - \pi/2 -\varphi_{\vt}$, the amplitude $\cA^{(0)}_n$ is expressed in terms of the integral $I_n(\rho)$:
\begin{eqnarray}
&& \cA^{(0)}_n = i^n e^{in\varphi_{\vt}}I_n(\rho),
\nonumber \\
&& I_n(\rho) = \frac{1}{2\pi}\int_{-\pi}^{\pi}f(x)e^{i\varphi(\rho,x)}dx,
\label{A:int}
\end{eqnarray}
where $f(x)=\cA(x+\pi/2+\varphi_{\vt})$ is a periodic function of $x$ and $\varphi(\rho,x) = nx - \rho\sin x$. Assuming $\rho\gg 1$
and $\rho\geq |n|$, the main contribution to the integral $I_n(\rho)$ is given by the neighborhoods of the saddle points $x=x_{\pm}$,
satisfying the equation $d\varphi(x)/dx = 0$:
\begin{equation}
\label{A:xpm}
x_{\pm} = \pm\alpha, \quad \cos\alpha=\frac{n}{\rho}, \quad 0\leq\alpha\leq\pi.
\end{equation}
Since the points $x_{\pm}$ tend toward each other and coalesce at $\alpha=0$, following the general idea of the uniform approximations of integrals \cite{Wong}, we rewrite the pre-exponential function $f(x)$, explicitly extracting the term, which approximates the $f(x)$ in the neighborhood of the two coalescing saddle points. Taking into account the periodicity of $f(x)$, we rewrite it in the following form:
\begin{equation}
\label{A:f}
f(x) = a_0 + a_1\sin x + (\cos x - \cos\alpha)g(x),
\end{equation}
where $a_0$ and $a_1$ are easily determined to be
$$
a_0 = \frac{f(x_{+})+f(x_{-})}{2},\quad a_1 = \frac{f(x_{+})-f(x_{-})}{2\sin\alpha},
$$
and where $g(x)$ is an analytic, smooth, periodic function of $x$. After substituting Eq.~(\ref{A:f}) into Eq.~(\ref{A:int}), the integration of the first two terms of the expression~(\ref{A:f}) can be performed analytically. The result for $I_n(\rho)$ is:
\begin{equation}
\label{A:result}
I_n(\rho) = a_0J_n(\rho) + ia_1 J_n'(\rho) + \tilde{I}_n(\rho),
\end{equation}
where $J_n(\rho)$ and $J_n'(\rho)$ are the Bessel function and its derivative, while $\tilde{I}_n(\rho)$ is the remainder integral:
\begin{equation}
\label{A:remainder1}
\tilde{I}_n(\rho) = \frac{1}{2\pi}\int_{-\pi}^{\pi}(\cos x - \cos\alpha)g(x)e^{i\varphi(\rho,x)}dx.
\end{equation}
Integrating $\tilde{I}_n(\rho)$ by parts, we obtain
\begin{equation}
\label{A:remainder2}
\tilde{I}_n(\rho) = \frac{1}{2\pi i\rho}\int_{-\pi}^{\pi}\frac{dg(x)}{dx}e^{i\varphi(\rho,x)}dx.
\end{equation}
Comparing Eq.~(\ref{A:remainder2}) with Eq.~\eqref{A:int}, one sees that the remainder term $\tilde{I}_n(\rho)$ has the same form as the original integral~(\ref{A:int}), but contains a small parameter $\rho^{-1}$. Representing the function $dg(x)/dx$ in Eq.~(\ref{A:remainder2}) by the form (\ref{A:f}) and applying the same integration procedure as for $I_n(\rho)$, we find the asymptotic expansion of the integral $I_n(\rho)$ for the large parameter~$\rho$.

For the case of a Kroll-Watson-like approximation,
we neglect the remainder term $\tilde{I}_n(\rho)$ in Eq.~(\ref{A:result}),
which gives immediately the result (\ref{ampl0}) for the scattering amplitude~$\cA^{(0)}_n$.

Also, we recall here another asymptotic approximation of the integral~(\ref{A:int}),
which was suggested in Ref.~\cite{Taulbjerg98}, where the integration interval in~Eq.~(\ref{A:int})
was divided into two parts ($-\pi\geq x\geq 0$ and $0\geq x\geq \pi$) followed by taking into
account the saddle points $x_{\pm}$ independently (as non-coalescing saddle points).
The result is that the integral $I_n(\rho)$ can be expressed in terms of the Anger
function, $J_n(\rho)$, (which coincides with the Bessel function for integer $n$) and the
Weber function, $E_n(\rho)$~\cite{abramovitz}:
\begin{eqnarray}
\label{A:Weber}
I_n(\rho)= a_{+} J_n(\rho) + ia_{-} E_n(\rho), \\
\nonumber
a_{\pm} = \frac{ f(x_{+}) \pm f(x_{-}) }{2}.
\end{eqnarray}



\begin{thebibliography}{12}
\bibitem{comm} Note that TDER theory, used in this paper, can be easily generalized to account for
two (e.g., $s$ and $p$) phase shifts, as was done in the TDER theory for bound state problems in
Ref.~\cite{two-stateTDER}.
\end{thebibliography}
\end{document}